\documentclass[journal]{IEEEtran}

\usepackage{float} 
\usepackage{tabularx}
\usepackage{mathtools}
\usepackage{amsmath}
\usepackage{hhline}
\usepackage{silence}
\usepackage{graphicx}
\usepackage{multirow} 
\usepackage{graphicx}
\usepackage{multirow}
\usepackage{amsmath}
\usepackage[english]{babel}
\usepackage{booktabs}
\usepackage{array}
\usepackage{paralist}
\usepackage{threeparttable}
\usepackage{lipsum}
\usepackage{flushend}
\usepackage{cuted}
\usepackage{algpseudocode}
\usepackage{algorithm}
\usepackage{algcompatible}
\usepackage{amssymb}
\usepackage{color}
\usepackage{psfrag}
\usepackage{epsfig}
\usepackage{multicol}
\usepackage{color}
\usepackage{fancybox}
\usepackage{subfigure}
\usepackage{cite}
\usepackage{theorem}
\usepackage{url}
\usepackage[table]{xcolor}
\usepackage{epstopdf}
\usepackage[normalem]{ulem}
\usepackage{siunitx}
\usepackage{rotating}
\usepackage{textcase}
\usepackage{dirtytalk}

\usepackage[justification=centering,font=small,labelfont=bf]{caption}
\usepackage{silence}
\usepackage{enumitem}
\definecolor{gray1}{gray}{0.90}
\definecolor{gray2}{gray}{0.98}
\definecolor{light-gray}{gray}{0.95}

\newcommand{\ignore}[1]{}
\newcommand{\redHL}[1]{\textcolor{red}{#1}}

\usepackage{tikz}
\usetikzlibrary{tikzmark}
\renewcommand{\baselinestretch}{0.93}

\begin{document}

\title{Exploring the Feasibility of Using 3D XPoint\\
		 as an In-Memory Computing Accelerator}
\author{Masoud Zabihi, Salonik Resch, H\"{u}srev C{\i}lasun, Zamshed I.
	Chowdhury, Zhengyang Zhao, \\
  	Ulya R. Karpuzcu, Jian-Ping Wang, and Sachin S.  Sapatnekar\thanks{The authors
	are with the Department of Electrical and Computer Engineering, University of
	Minnesota, Minneapolis, MN, USA. This work was supported in part by NSF SPX
	Award CCF-1725420.}
}

\maketitle

\begin{abstract}
This paper describes how 3D XPoint memory arrays can be used as in-memory computing accelerators. We first show that thresholded matrix-vector multiplication (TMVM), the fundamental computational kernel in many applications including machine learning, can be implemented within a 3D XPoint array without requiring data to leave the array for processing. Using the implementation of TMVM, we then discuss the implementation of a binary neural inference engine. We discuss the application of the core concept to address issues such as system scalability, where we connect multiple 3D XPoint arrays, and power integrity, where we analyze the parasitic effects of metal lines on noise margins. To assure power integrity within the 3D XPoint array during this implementation, we carefully analyze the parasitic effects of metal lines on the accuracy of the implementations. We quantify the impact of parasitics on limiting the size and configuration of a 3D XPoint array, and estimate the maximum acceptable size of a 3D XPoint subarray.     
\end{abstract}	

\section{Introduction} \label{sec:intro}

\noindent

With the rapidly increasing sizes of datasets and challenges in transistor scaling in recent years, the need for new computing paradigms is felt more than ever~\cite{bigData}. In today's computing systems, large portions of computation energy and time are wasted for transferring data back and forth between the processor and the memory~\cite{Keckler}. One approach is to bring the processor and memory closer to each other and build a near-memory platform that places the computing engine adjacent to the memory, and hence reduce the energy and time overhead for data transfer. Another approach that even more significantly reduces the time overhead and energy is to use the memory device as the computational unit and built a {\em true in-memory} computing platform. We follow the latter approach. 

The substrate that we work on is 3D XPoint~\cite{Intel_talk}, a class of memory technology that fills a unique place within the memory hierarchy between solid state storage drive (SSD) and the system main memory. In comparison with the NAND-based SSD (which is the most ubiquitous storage device available today~\cite{SSD_design_tradeoffs}), it has the advantage of being faster, denser, and more scalable. Its nonvolatility differentiates it from competing technologies such as NAND-based SSDs and dynamic random access memories (DRAMs), although NAND-based SSDs are more cost-effective today and DRAMs are faster. 3D XPoint is fabricated using monolithic 3D integration of a nonavolatile memory stack on top of CMOS peripheral circuitry.

The operation and performance of 3D XPoint as a memory unit are discussed in~\cite{KIST_2020, KIST_2019, KIST_2018,3DICT, Lilja_3DXpoint}. In our work, rather than focusing again on the memory aspects of 3D XPoint, we explore the possibility of exploiting 3D XPoint arrays to perform in-memory computation. This means not only that 3D XPoint can function as a storage unit, but also that it can perform computation inside its array without the need for the data to leave the array. Therefore, unlike conventional computational systems, the information can be processed locally rather than being sent to a processor through the memory hierarchy. The analysis in this paper considers wire non-idealities and physical design of 3D XPoint subarray. We first show the implementation of thresholded matrix-vector multiplication (TMVM), which is a building block for neural networks (NNs) and deep learning applications. Second, using this core operation, we discuss the implementation of a neural network inference engine. Finally, we discuss how to enable 3D XPoint for more complex versions of these implementations (e.g. multi-bit operations and multi-layer NNs).

For in-memory computing platforms, wire resistances are a substantial source of non-ideality that must be taken into account during the implementations~\cite{parasitics_importance_2020}. In ~\cite{CRAM_parasitics}, an analytical approach is developed to study the effects of the parasitic of wires for the implementation on spintronics computational RAM. In this work, we develop a comprehensive method to analyze the impact of wire parasitics of wires in the 3D XPoint subarray. We devise a methodology that helps us to determine the maximum acceptable size of a 3D XPoint subarray that ensures electrically correct operation of the computation. Our method considers various parameters (such as width, length, and configurations of metal lines) to calculate the wire parasitics and incorporates them in the implementation.  

Next, we discuss the structure of 3D XPoint in Section~\ref{sec:background}. In Section~\ref{sec:realization}, we describe the implementation of TMVM, and NN. In Section~\ref{sec:more_complex}, we explore the methods for more complex implementations. We develop the models for the effect of wire parasitics in Section~\ref{sec:parasitics}, evaluate the results of our analysis in Section~\ref{sec:results and Discussion}, and then conclude the paper in Section~\ref{sec:conclusion}.

\section{Overview of 3D XPoint}
\label{sec:background}
\begin{figure}[ht!]
	\centering
	\includegraphics[width=8.0cm, height=4.0cm]{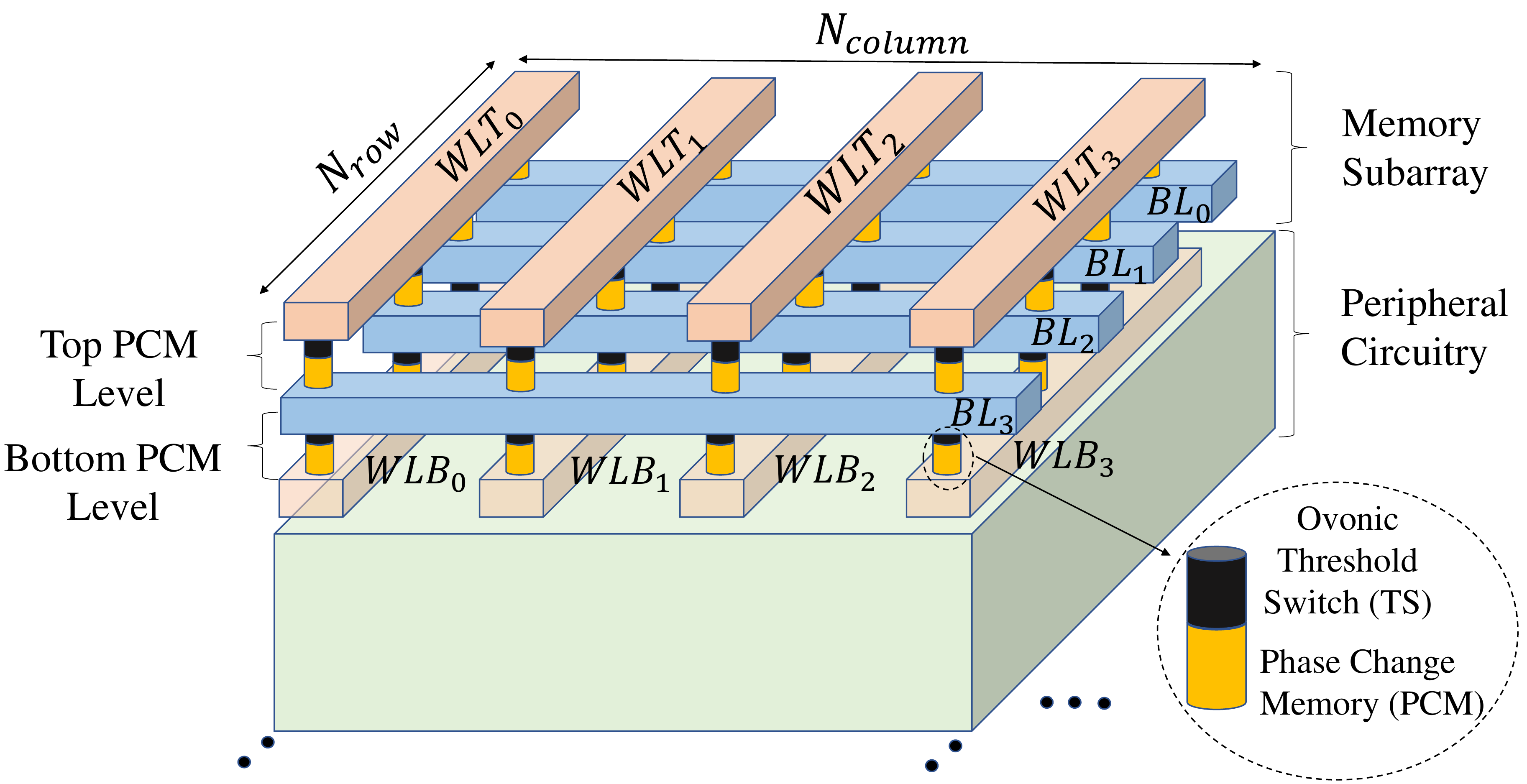}
	\caption{The structure of a 3D XPoint subarray. The CMOS peripheral circuitry is located underneath the memory subarray.}
	\label{fig:3DXpoint_overview}
\end{figure}

\noindent
Fig.~\ref{fig:3DXpoint_overview} shows the overall structure of a 3D XPoint subarray. A two-level PCM stack is integrated at the top of CMOS peripheral circuitry. The storage device is based on phase-change memory (PCM) technology, which is connected to a compatible ovonic threshold switch (OTS) made of AsTeGeSiN~\cite{compatible_OTS, Burr2014AccessD, stackable_3D_PCM}. Word lines at the top ($WLT$s), word lines at the bottom ($WLB$s), and bit lines ($BL$s) in the middle provide the current path to each individual memory cell~\cite{OTS-PCM}. The compatibility of the junction of PCM and OTS devices is a key factor in allowing access to individual cells without facing sneak path problems~\cite{sneak_path}. The total number of PCM cells in the 3D XPoint subarray with $N_{row}$ rows and $N_{column}$ columns is $\left(2 N_{row} \times N_{column}\right)$, with half in the top PCM level and half in the bottom PCM level, as shown in the figure.

PCM is a non-volatile memory technology exploiting Ge–Sb–Te (GST) alloys (e.g., Ge$_2$Sb$_2$Te$_5$) as the storage medium~\cite{GST_2011}. PCM has two states: a crystalline phase with high conductance ($G_{C}$) and an amorphous phase with low conductance ($G_{A}$). The GST alloy transition between amorphous and crystalline states is triggered by changing the temperature level ~\cite{PCM_temprature_2010, review_2016}. In early explorations of PCM technology (1970s--early 2000s), the temperature level was changed using a laser source~\cite{optical_PCM}. The state-of-the-art research on PCM is focused on using electrical impulses to change the temperature, and hence the state, of the PCM device by applying an electric current (or voltage) pulse across the PCM device~\cite{review_Zheng_2017}.

\begin{figure}[ht!]
	\centering
	\subfigure[]{\includegraphics[width=6.5cm, height=3cm]{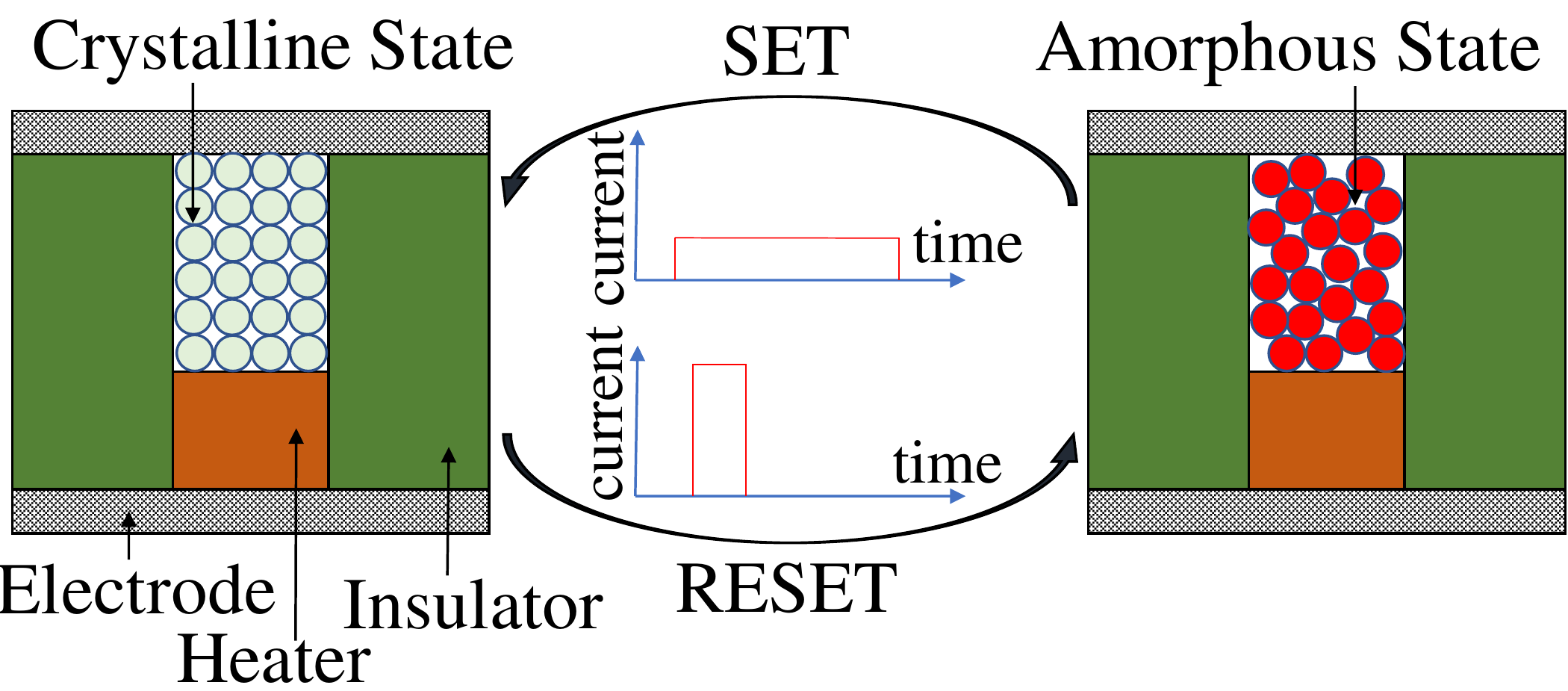}}
	\subfigure[]{\includegraphics[width=2cm, height=3cm]{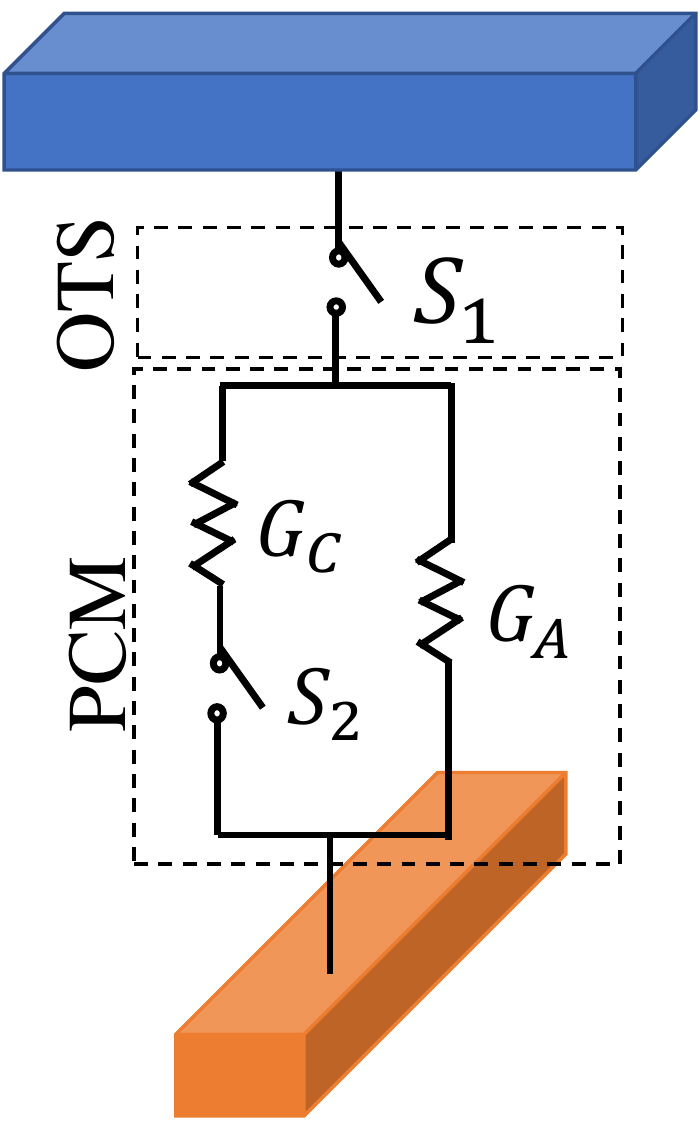}}
	\caption{PCM model: (a) the transition between amorphous and crystalline phases by applying SET and RESET pulses across a pillar type PCM device, and (b) PCM cell can be modeled using  a resistive circuit with two voltage control switches~\cite{KIST_2018}.}
	\label{fig:PCM_model}
\end{figure}

Fig.~\ref{fig:PCM_model}(a) shows that applying a fast high-amplitude current pulse of amplitude $I_{RESET}$ (called the RESET pulse) heats up the GST material to the melting temperature $T_{melt}$ ($\sim$600$^{\circ}$C or higher ~\cite{review_2016}), erasing the previous periodic and ordered atomic arrangement. After quenching, the new disordered atomic structure will be frozen, making the transition from high conductance crystalline state to low conductance amorphous state possible. To change the state of the GTS from amorphous to crystalline, a slow, relatively low amplitude current pulse of amplitude $I_{SET}$ (called the SET pulse) must pass through the GST material. The SET current pulse causes the GST material to heat up to crystalline temperature $T_{cryst}$ ($\sim$400$^{\circ}$C~\cite{review_2016}). Over a long SET time of several tens of nanoseconds, this is a high enough temperature (still lower than $T_{melt}$) for the reconfiguration and crystallization of the previous amorphous atomic region to the crystalline state. The desirable PCM characteristics are a lower amplitude of the RESET current and a shorter SET time. A RESET current as low as 10$\mu$A and a SET time as low as 25ns for individual PCM devices is already demonstrated with sub-20nm scalability, high endurance 10$^{12}$ cycles, and a projected 10-year retention time at 210$^{\circ}$C~\cite{review_2020}.

\ignore{Other important parameters of PCM are high endurance and retention time (more than $10^{8}$ RESET/SET cycle and more than 10 years at 85$^{\circ}$C, respectively ~\cite{review_2020}).} 

Fig.~\ref{fig:PCM_model}(b) shows the electrical model of PCM cell. The resistance across the PCM cell can be modeled by two voltage controlled switches~\cite{KIST_2018}. Depending on the status of switches $S_{1}$ and $S_{2}$, different currents flow between two lines connected to the terminals of the PCM cell, determined by $G_{A}$ and $G_{C}$. The ON/OFF states of the memory cell are determined by OTS: If the voltage level across the OTS of a cell is larger than a threshold, the cell is considered to be ON, and it is OFF otherwise. In today's technologies, the OTS conductance for the OFF state is up to $10^8\times$ smaller than for the ON state.

The value stored in the PCM device can represent either logic 1 (crystalline phase) or logic 0 (amorphous phase). Three memory operations available in 3D XPoint: write logic 1 (using the fast high-amplitude SET pulse), write logic 0 (using long low-amplitude RESET pulse), and read. For the memory read operation, since it is undesirable to change the state of the PCM cell, a pulse with relatively very small amplitude will be applied, increasing the temperature slightly above the ambient temperature but below $T_{cryst}$ (and of course $T_{melt}$).

\section{Realization of In-Memory Computing}
\label{sec:realization}

\subsection{Implementation of TMVM}
\noindent
TMVM is a fundamental step in the implementation of many applications, and is a fundamental computational kernel in machine learning (ML) applications. Using 3D XPoint as the TMVM computation engine can tremendously decrease the ML computational workload, as the data does not need to leave the 3D XPoint array during the computation. 

To show how the first step of a TMVM, let us multiply, without thresholding, matrix $G \in \mathbb{R}^{(N_{x}+1) \times (N_{y}+1)}$ and vector $V=\left[V_{0} V_{1} V_{2} ... V_{N_{x}}\right]^T \in \mathbb{R}^{(N_x+1)}$, where $G$ is given by:
\begin{equation}
G=
\begin{pmatrix}
G_{0,0} & G_{0,1} & ...&  G_{0,N_{y}}\\
G_{1,0} & G_{1,1} & ... &  G_{1,N_{y}}\\
    .    &   .       & ... &    .\\
    .    &   .       & ... &    .\\
G_{N_{x},0} & G_{N_{x},1} & .. & G_{N_{x},N_{y}}\\
\end{pmatrix}
\label{eq:G}
\end{equation}
This computes $O=\left[O_{0} O_{1} O_{2}, ... O_{N_{y}}\right]^T \in \mathbb{R}^{(N_{y}+1)}$) where each element of vector $O$ is a dot product. For example,
\begin{equation}
O_{0}=G_{0,0}V_{0}+G_{1,0}V_{1}+...+G_{N_{x},0}V_{N_{x}}
\label{eq:O_0}
\end{equation}
This is computed in the 3D XPoint array by applying voltages across a set of conductances to produce a current $O_0$.

Today's PCM cells can only store binary values. Hence, we assume that elements of matrix $G$ and vector $V$ are binary. To implement a \say{neuron-like} operation using TMVM, the $O_{0}$ value and computed $O_{i}$ values are followed a thresholding operation. In~\eqref{eq:O_0}, if the sum of products exceeds the current required to flip the output bit, then logic 1 is stored as the conductance, $G_{O_0}$, of the PCM cell $O_0$; otherwise, the stored logic value is 0. Similarly, for other $O_{i}$s, the values after thresholding are stored as the conductance states, $G_{O_i}$.

\begin{figure}[ht!]
	\centering
	\includegraphics[width=9.0cm, height=3.0cm]{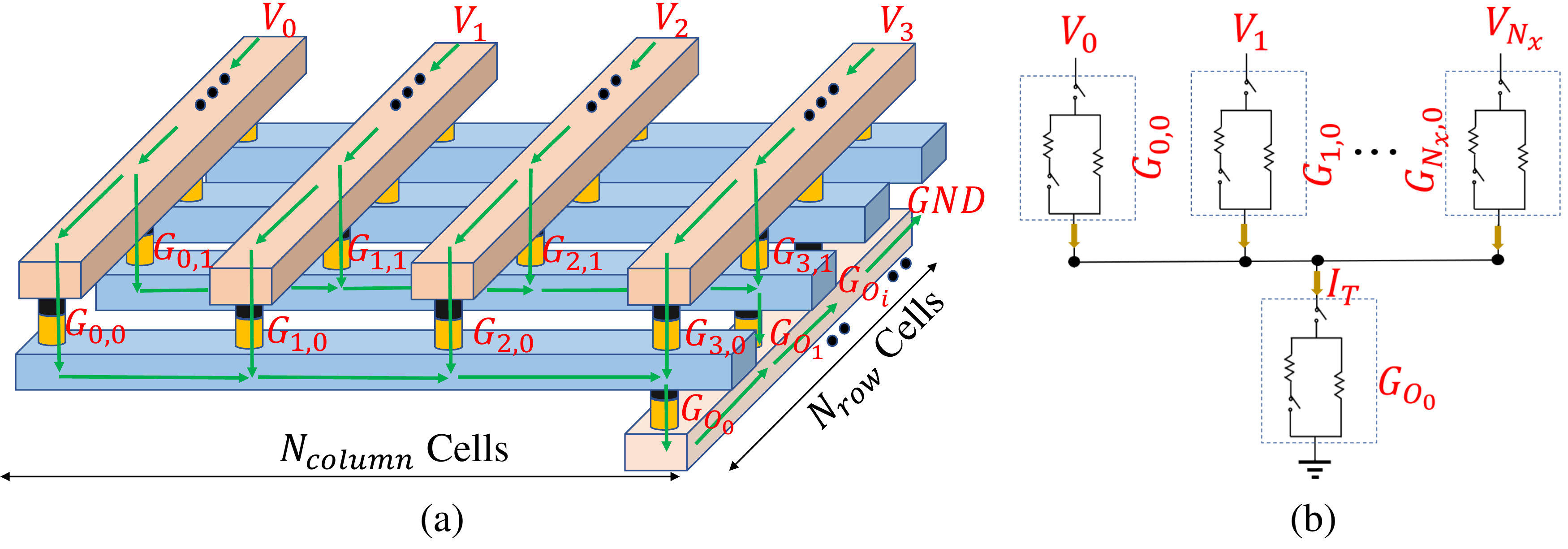}
	\caption{(a) Using 3D XPoint as an in-memory computing engine for TMVM of $GV$. (b) The equivalent circuit model for the implementation of a dot product (to calculate $O_{0}$).}
	\label{fig:3D XPoint_dot_product}
\end{figure}

Fig.~\ref{fig:3D XPoint_dot_product}(a) shows the implementation of the TMVM on a 3D XPoint subarray with $\left(2 N_{row} \times N_{column}\right)$ PCM cells ($\left(N_{row} \times N_{column}\right)$ cells each at top PCM level and bottom PCM level) where $N_{column}=N_{x}+1$ and $N_{row}=N_{y}+1$. For clarity, as compared to Fig.~\ref{fig:3DXpoint_overview}, only the lines and PCM cells engaged in the computation are shown, and the rest of the lines and the PCM cells (at the bottom) are removed from the figure. All elements of $O$ will be calculated simultaneously and are stored in the same column with $N_{row}$ PCM cells. Considering that today's 3D XPoint cannot store multiple values in a cell, we assume that elements of vector $V$ and $G$ are binary.

The conductances $G$ are first programmed in the top PCM level by memory write operations or by previous computation.

\begin{itemize}

\item Before the computation starts, cells that store $G_{O_{i}}$s at the bottom are preset to logic 0.
\item Then, voltages $V_i, 0 \leq i \leq N_x$ are applied to the word lines $WLT$s connected to input cells located at top. If $V_{i}$ represents logic 1, voltage $V_{DD}$ is applied ($V_i \leftarrow V_{DD}$) to the $WLT_{i}$ and the current that flows through the corresponding input cell is proportional to $G_{0,i}V_{DD}$.
\item If $V_{i}$ represents logic 0, $WLT_{i}$ is floated ($V_i \leftarrow$ {\em float}) and no current passes through the corresponding PCM cell. 
\item The summation of currents ($I_{T}$) from input cells flows through the $G_{O_{0}}$ in a time interval $t_{SET}$. Based on the values of $V_{i}$ and $G_{i,0}$, different currents pass through the input cells that store $G_{i,0}$. If $I_T > I_{SET}$, the state of $G_{O_{0}}$ changes to logic 1. However, we require $I_T <I_{RESET}$ to avoid erroneous computation. 

\end{itemize}

To calculate the minimum and maximum allowable applied voltage ($V_{DD}$) to the lines, we consider a simplified electrical model for the implementation of a dot product (e.g., for $O_{0}$) shown in Fig.~\ref{fig:3D XPoint_dot_product}(b). Current $I_{T}$ can be written as follows
\begin{equation}
I_{T}=G_{O_{0}}\frac{\sum_{i=0}^{N_{x}} V_{i}G_{i,0}}{\sum_{i=0}^{N_{x}}G_{i,0}+G_{O_{0}}}
\label{eq:I_T}
\end{equation}
When the computation begins, $G_{O_{0}}~\approx~G_{A}$ since the preset is 0, and $I_{T}(t=0)$ is small (of the order of few hundred nAs). However, by the passage of time, the amorphous region near the heater in the output PCM starts to turn crystalline, resulting in increasing $G_{O_{0}}$ (and consequently $I_{T}$) and heat (generated by the flow of more electric current). If the applied voltage $V_{DD}$ is large enough to provide a current larger than $I_{SET}$, crystallization repeats until a threshold point where the whole amorphous region in the output PCM turns into a crystalline region with high conductivity, representing logic 1. On the other hand, the $V_{DD}$ must not be so large that the generated temperature exceeds $T_{melt}$, causing erroneous computation.

To calculate the $V_{DD}$ range for the accurate implementation of the described dot product, we analyze the two cases corresponding to $V_{min}$ (the minimum acceptable voltage) and $V_{max}$ (the maximum acceptable voltage). For the $V_{min}$ case, we assume that all $V_{i}$s, and all $G_{i,0}$s represent logic 1, i.e., $V_{DD}$ voltages are applied to all $WLT$s and the conductances of input cells are in the high conductance state corresponding to $G_{C}$. In this case, from~\eqref{eq:I_T}, $I_{T}=\left(\frac{N_{x}+1}{N_{x}+2}\right)G_{C}V_{DD}$. Since $I_{SET}\leq I_{T} \leq I_{RESET}$, the $V_{min}$ requirement implies a first constraint, requiring that $V_{DD}$ to lie in the range:
\begin{equation}
{\cal R}_1=\left[\left(\frac{N_{x}+2}{N_{x}+1}\right)\left(\frac{I_{SET}}{G_{C}}\right), \left(\frac{N_{x}+2}{N_{x}+1}\right)\left(\frac{I_{RESET}}{G_{C}}\right)\right] 
\label{eq:R1_constraint}
\end{equation}

For the $V_{max}$ case, all $V_{i}$s are set to logic 1, while all $G_{i,0}$s are set to logic 0. Since the result of the dot product should be at logic 0, we expect that the preset value of PCM stored $O_{0}$ remains intact. At the maximum voltage possible, we hypothetically assume that the applied voltage pulse should be below the level required to change the output state from logic 0 to logic 1, even if conductances of all input cells are $G_{A}$. In other words,  $I_{T}=\left(\frac{(N_{x}+1)G_{A}G_{C}}{(N_{x}+1)G_{A}+G_{C}}\right)V_{DD}<I_{SET}$, i.e., the output state cannot be altered. Therefore, the second set of constraints require $V_{DD}$ to lie in the range
\begin{equation}
{\cal R}_2=\left[0, \left(\frac{(N_{x}+1)G_{A}+G_{C}}{(N_{x}+1)G_{A}G_{C}}\right)I_{SET}\right] 
\label{eq:R2_constraint}
\end{equation}
The acceptable range for $V_{DD}$ is ${\cal R}_1 \cap {\cal R}_2$. Therefore, the minimum and maximum acceptable voltages are $V_{min}=\min({\cal R}_1)$ and $V_{max}=\min(\max({\cal R}_1), \max({\cal R}_2))$, respectively.

\ignore{                
\begin{equation}
\begin{aligned}
 V_{DD}& \in R_{1} \cap R_{2}=\left[\frac{(N_{x}+2)I_{SET}}{(N_{x}+1)G_{high}},\\
 &\min(\frac{(N_{x}+2)I_{RESET}}{(N_{x}+1)G_{high}},(\frac{1}{G_{high}}+\frac{1}{(N_{x}+1)G_{low}})I_{SET})\right]
\end{aligned} 
\label{eq:R1_R2_intersection}
\end{equation}
}

\subsection{Implementation of NN}

\begin{figure}[ht!]
	\centering
	\includegraphics[width=8.0cm, height=5.5cm]{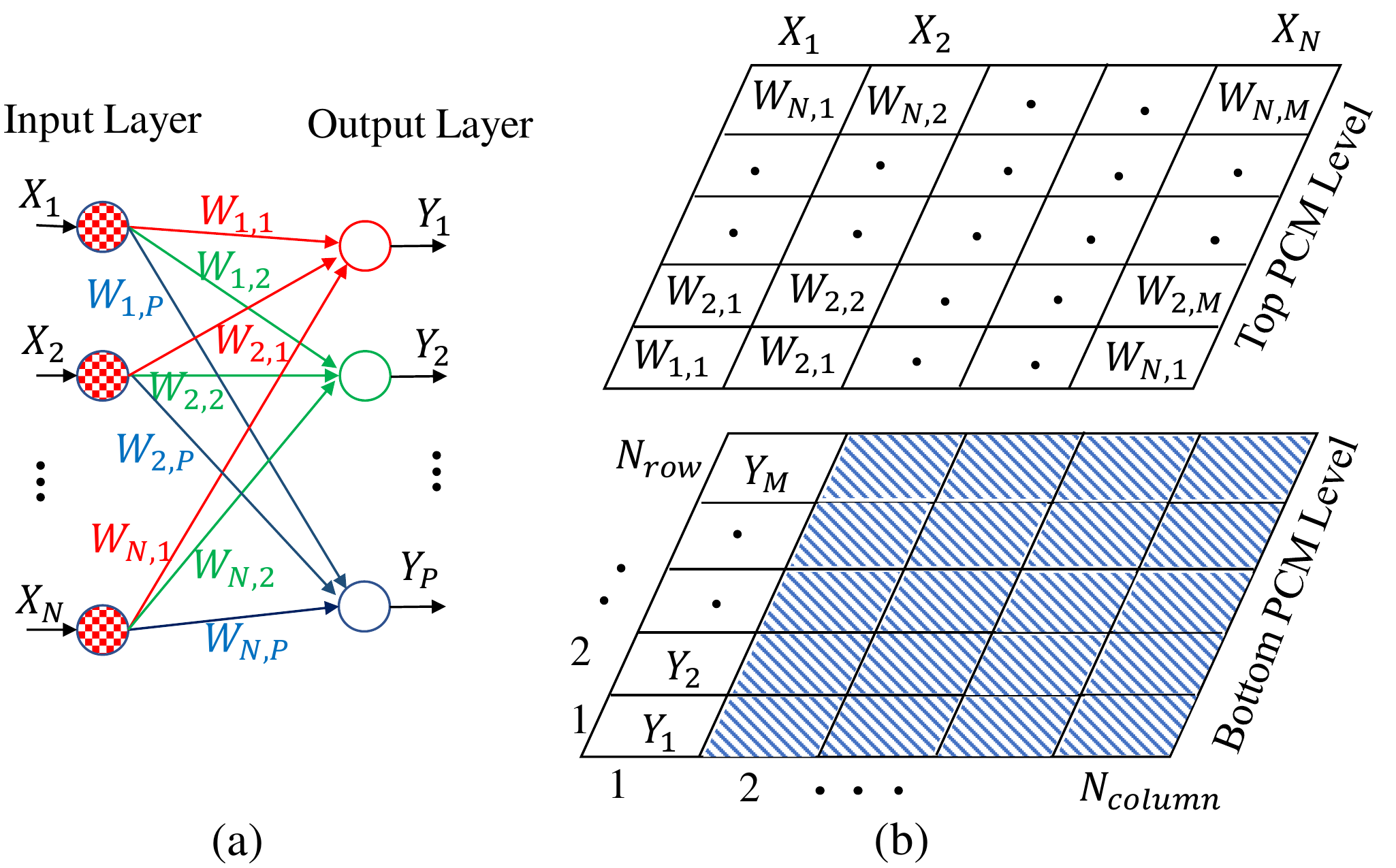}
	\caption{NN implementation: (a) A single-layer neuromorphic
		inference engine. (b) Data layout for the NN implementation.}
	\label{fig:NN}
\end{figure}

\noindent
Using the TMVM implementation, we implement a neuromorphic inference engine. Fig.~\ref{fig:NN}(a) shows a single-layer NN with $N$ inputs and $P$ outputs. Fig.~\ref{fig:NN}(b) shows the data layout for the NN implementation on a 3D XPoint subarray. The top PCM cells are allocated for storing the weights ($W_{i,j}$s), similar to $G_{i,j}$s in TMVM that were stored in the top PCM cells, and the bottom PCM cells are allocated for storing the outputs ($Y_{i}$s), similar to $O_{i}s$ in TMVM. The inputs ($X_{i}$s) are applied to $WLT_{i}$s as voltage pulses (similar to $V_{i}$s in TMVM).If $N \leq N_{column}$ and $P \leq N_{row}$, then all $Y_{j}$s can be determined simultaneously in one step. The output elements of the NN can be stored in any column at the bottom (here, we choose column 1). In Fig.~\ref{fig:NN}(b), all other cells at the bottom patterned by diagonal stripes are not engaged in the computation of $Y_{i}$s, i.e., $WLB$s connected to these cells are floated.     

An application for the proposed NN implementation is handwritten digit recognition of MNIST dataset with 10K test images~\cite{MNIST}. Analyzing each MNIST test image can be performed using a similar NN shown in Fig.~\ref{fig:NN}(a). Here $P=10$, as in MNIST each image represents a digit (from 0 to 9). In each computational step, $\lfloor{\frac{N_{row}}{P}}\rfloor$ images can be processed and stored in a column.

\section{Enabling More Complex Implementations}
\label{sec:more_complex}
\noindent
In this section, we discuss three concepts that enable more complex computations. Then, we provide the implementation of a multi-layer NN as an example.  

\subsection{3D XPoint with four stacked level of PCM cells} 
\begin{figure}[ht!]
	\centering
	\includegraphics[width=5.0cm, height=3cm]{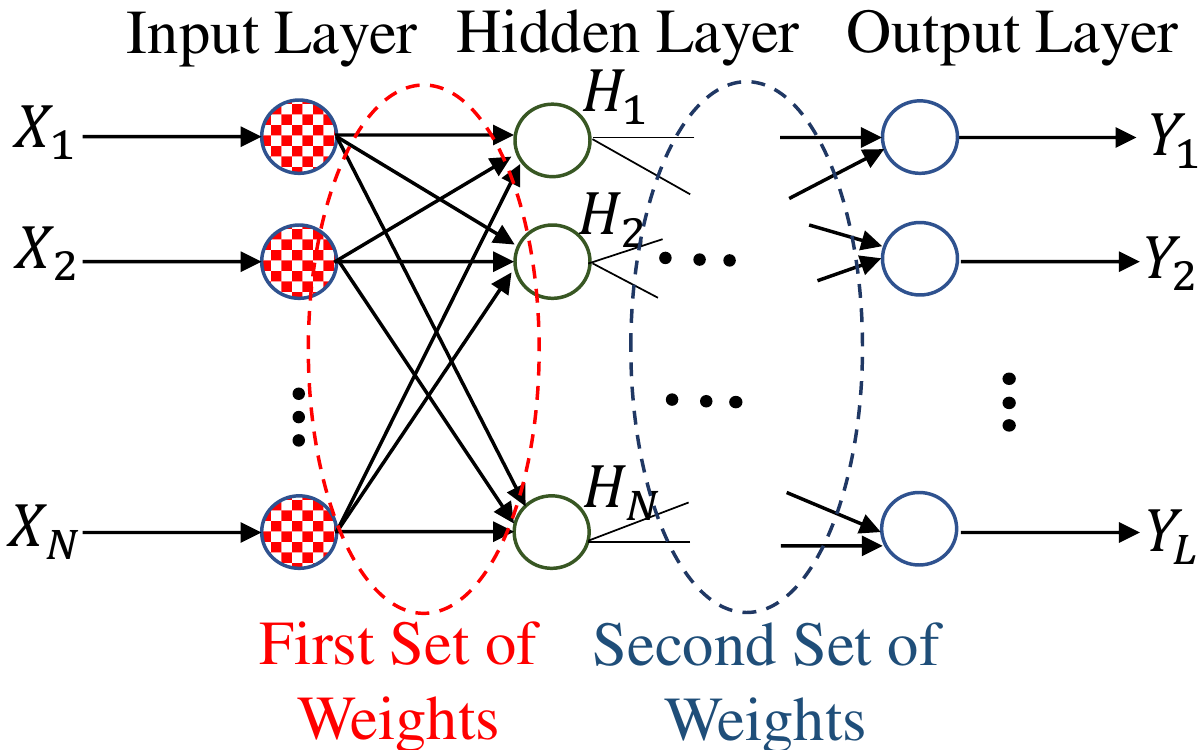}
	\caption{Multi-layer NN with an input, hidden, and output layer.}
	\label{fig:multi_layer_NN}
\end{figure}

\noindent
Industry projections show that the next generation of 3D XPoint will have four-level stacked PCM cells~\cite{4-level-stack}. If the number of PCM levels increases, then the volume of stored information per footprint area increases, and more complex implementations are possible.  Although a two-level subarray of PCM cells is sufficient to implement any NN (see Section~\ref{sec:more_complex}D), we  will illustrate how we can use a four-level subarray of PCM cells to implement a multi-layer NN by exploiting the extra PCM levels.  The NN in Fig.~\ref{fig:multi_layer_NN} has three layers: an input layer, a hidden layer, and an output layer. At the top PCM level, the first set of weights are stored. In the next PCM level, the hidden layer data is calculated, and by applying the second set of weights as voltage pulses, we obtain the outputs $Y_i$ of the NN at the third PCM level.    

\subsection{Scalability of 3D XPoint to large computations}

\begin{figure}[ht!]
	\centering
	\includegraphics[width=9.0cm, height=6.0cm]{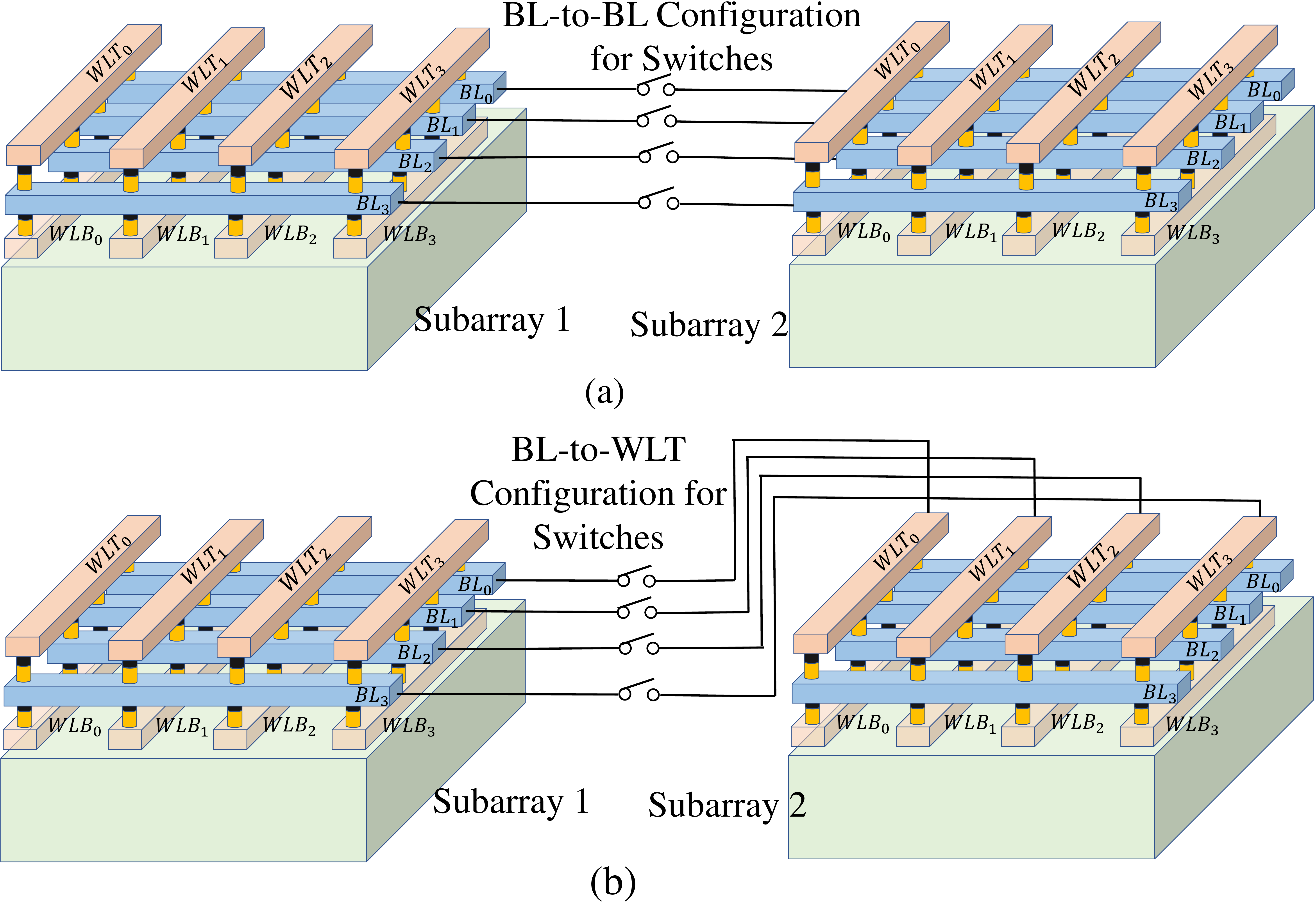}
	\caption{Two configurations for communication between 3D XPoint subarrays: (a) switches connect $BL$s of subarray 1 to $BLs$ of Subarray 2, and (b) switches connect $BL$s of the subarray 1 to $WLT$s of subarray 2.}
	\label{fig:subarray_communication}
\end{figure}

\noindent
We can connect multiple 3D XPoint subarrays to create a larger array to handle computations with higher matrix dimensions. In Fig.~\ref{fig:subarray_communication}(a), switches connect $BL$s of subarray~1 to those of subarray~2, enabling current flow from the $BL$s of subarray~1 to those of subarray~2. The $WLB$ of subarray~2 that is  scheduled to store the computation results will be connected to ground, while all other $WLB$s not engaging in the computation (in both subarrays~1 and~2) are floated. Hence, the computation results in subarray~1, are being calculated and stored at the bottom PCM level of subarray~2 ($BL$-to-$WLT$). In Fig.~\ref{fig:subarray_communication}(b), switches connect $BL$s of the subarray~1 to $WLT$ of subarray~2. In this configuration, the results are being calculated at the top PCM level of subarray~2. The status of lines during the computation for these two configurations are listed in the Supplementary Materials.       
	
\subsection{Multi-bit operations}

\noindent	
Thus far, we have discussed the implementations of operations with binary digits. We introduce two methods that enable us to implement operations with multi-bit digits. For brevity, we explain the principle using two-bit digits, where each digit consists an $MSB$ (most significant bit) and an $LSB$ (least significant bit). Let us assume that we want to perform TMVM of $G_{2bit}V$ where $G_{2bit}$ is a $(N_{x}+1)\times(N_{y}+1)$ matrix with two-bit elements, meaning the element located in row $i$ and column $j$ of the matrix $G_{2bit}$ is $G_{i,j}^{MSB}G_{i,j}^{LSB}$.

\begin{figure}[ht!]
	\centering
	\includegraphics[width=9.0cm, height=2.0cm]{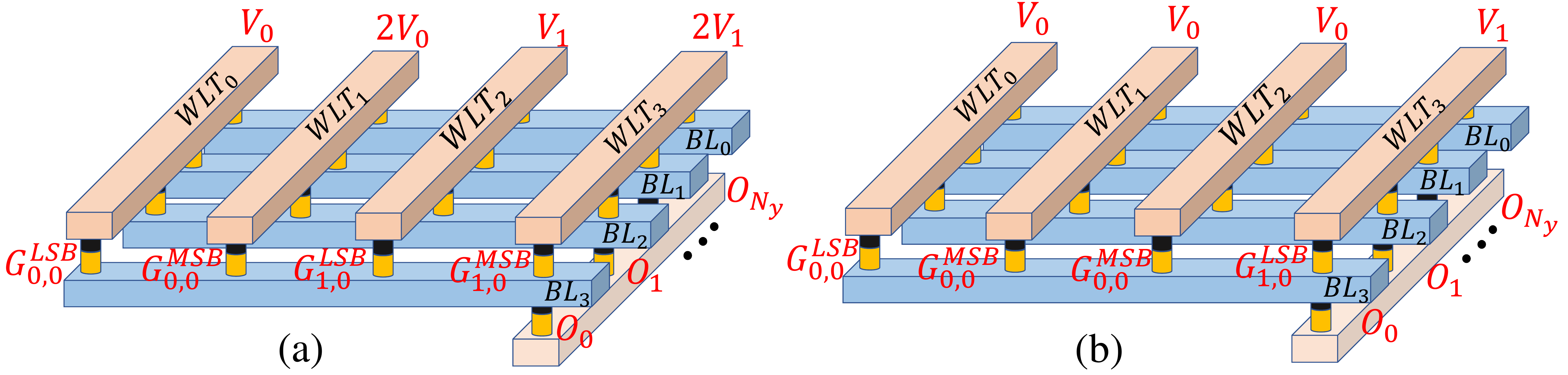}
	\caption{Two implementations with multi-bit operations: (a) area-efficient implementation, (b) low-power implementation.}
	\label{fig:multi-bits}
\end{figure}

Fig.~\ref{fig:multi-bits} illustrates two ways to implement 2-bit operations on a 3D XPoint subarray. Fig.~\ref{fig:multi-bits}(a) shows an area-efficient approach. For example, to calculate $O_{0}$, we need to calculate $(G_{0,0}^{LSB}+2G_{0,0}^{MSB})V_{0}+ (G_{1,0}^{LSB}+2G_{1,0}^{MSB})V_{1}+...+(G_{N_{x},0}^{LSB}+2G_{N_{x},0}^{MSB})V_{N_{x}}$. To do so, we can apply $V_{0}$ to the $WLT_{0}$ (connected to the PCM storing $G_{0,0}^{LSB}$ bit), and we can apply $2V_{0}$ to $WLT_{1}$ (connected to the PCM storing $G_{0,0}^{MSB}$ bit). Therefore, the current flowing through the $MSB$ cell is two times larger than that of the $LSB$ cell. Another more area-intensive approach, which does not require multiple voltage levels, is shown in Fig.~\ref{fig:multi-bits}(b) where we copy the $MSB$ in pair of adjacent cells, and we apply the same voltage to their corresponding $WLT$s. The current through the $MSB$ cell is weighted to be twice that of the current through the $LSB$ cell.  

\ignore{
 The  where the elements $G_{2bit}$ (see Equation~\ref{eq:G}) are 2-bits digits, the $the $

 if we want to replicate the implementation of $GV$ using whi With this in mind, if in the TMVM, the elements of $G$ are 2-bit digits, the matrix $G$ can be rewritten as follows 
	
	\begin{equation}
	G_{2-bit}=\\
	\begin{pmatrix}
	\overline{G_{0,0}^{MSB}G_{0,0}^{LSB}} & ...&  \overline{G_{0,N_{y}}^{MSB}G_{0,N_{y}}^{LSB}}\\
	\overline{G_{1,0}^{MSB}G_{1,0}^{LSB}} & ... & \overline{G_{1,N_{y}}^{MSB}G_{1,N_{y}}^{LSB}}\\
	.    &  ... &    .\\
	.    &   ... &    .\\
	\overline{G_{N_{x},0}^{MSB}G_{N_{x},0}^{LSB}} & .. & \overline{G_{N_{x},N_{y}}^{MSB}G_{N_{x},N_{y}}^{LSB}}\\
	\end{pmatrix}
	\label{eq:G_2bits}
	\end{equation}
	}
\ignore{
To calculate the TMVM of $G_{2-bits}$ and $V$, we can follow two approaches. One of the approaches is area efficient, that is, the required minimum number of PCM cells for the implementations. The other one is energy efficient, which is required minimum implementation energy without area considerations. 
}

\begin{figure}[ht!]
	\centering
	\includegraphics[width=9.0cm, height=6.cm]{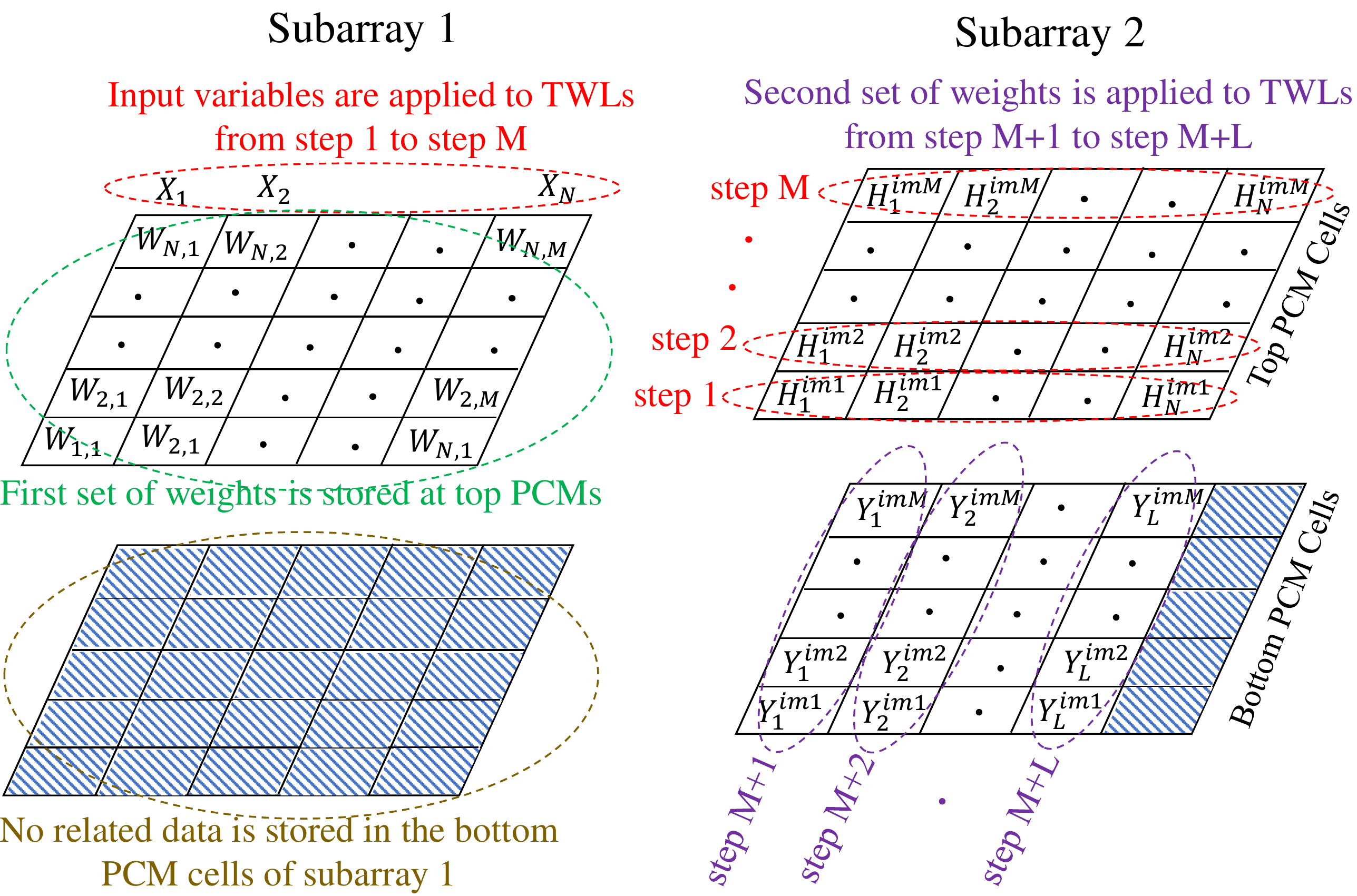}
	\caption{Data layout for the implementation of 3-layer NN. }
	\label{fig:data_layout}
\end{figure}
	
\subsection{Multi-layer NN implementation in a two-level PCM stack}

\noindent	
We now illustrate how a multi-layer NN can be implemented using a three-layer PCM stack. As an example, we discuss the implementation of the three-layer NN (shown in Fig.~\ref{fig:multi_layer_NN}) using two two-level 3D XPoint subarrays. Let us assume that the NN is required to analyze 10K images of the MNIST dataset. The data layout of this implementation is illustrated in Fig.~\ref{fig:data_layout} using two subarrays connected with $BL$-to-$WLT$ configuration (see Fig.~\ref{fig:subarray_communication}(b)). The first set of weights is stored at the top PCM cells of subarray 1. The inputs ($X_{0}, X_{1}, ..., X_{N}$) are applied as the voltages to the $WLT$s of subarray 1. We assume that at each time step, the hidden layer values ($H_{1}, H_{2}, ..., H_{N}$) for a specific image from MNIST are being processed. For example, the hidden layer values of the second image ($H_{1}^{im2}, H_{2}^{im2}, ..., H_{N}^{im2}$) is calculated in step 2. Assuming that we calculate the hidden layer values of $M$($=N_{row}$) images in each set of computation, we require $M$ steps to calculate and store the hidden layer values of $M$ images at the top PCM cells of subarray 2. In each of these steps, the corresponding $BL$ in subarray 2 is connected to $GND$, and the remaining $BL$s in subarray 2 are floated. After all hidden layer values are stored at the top PCM cells of subarray 2, we apply the second set of the weights (as voltage pulses) to the $WLT$s of subarray 2. At each column at the bottom PCM cells of subarray 2, the outputs ($Y_{i}$s) of $M$ images are calculated and stored.

\ignore{
Fig.~\ref{fig:multi_layer_NN} shows a multi-layer NN with an input layer, a hidden layer, and an output layer.      
}

\section{Analyzing Interconnect Parasitic Effects}
\label{sec:parasitics}

\begin{figure}[ht!]
 	\centering
 	\includegraphics[width=9cm, height=5.5cm]{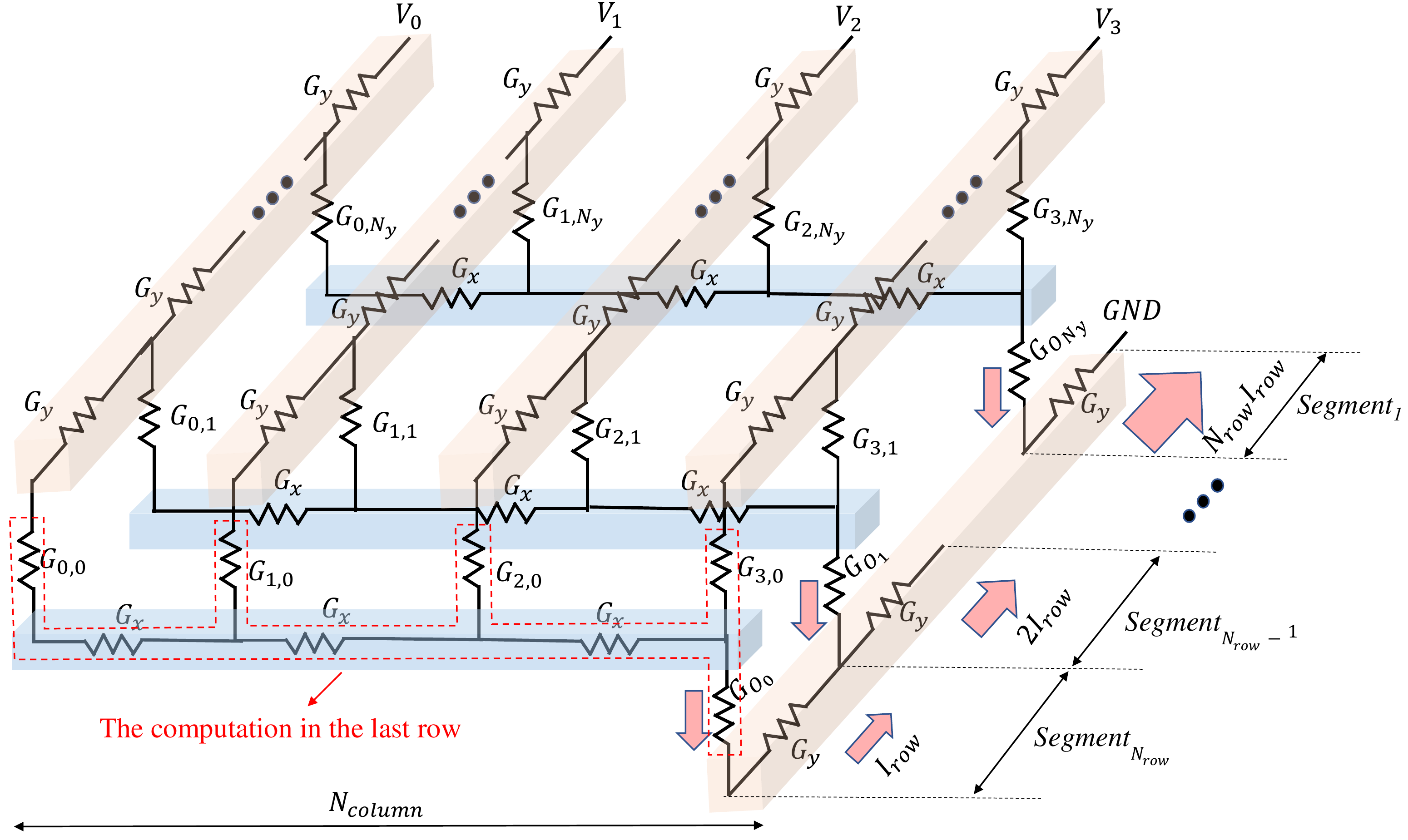}
 	\caption{The equivalent circuit model for the TMVM implementation with considering wire parasitics.}
 	\label{fig:parasitics_crk}
\end{figure}

\noindent
To ensure the electrical correctness of the implementations in in-memory computing platforms, we must consider non-idealities due to wire parasitic effects~\cite{parasitics_importance_2020,CRAM_parasitics}.  As an example, we consider the implementation of a TMVM illustrated in Fig.~\ref{fig:3D XPoint_dot_product}(a). In thhe equivalent circuit model shown in Fig.~\ref{fig:parasitics_crk}, the  $WLT$s, $BL$s, and $WLB$s have  nonzero parasitics that cause a voltage drop in the current path across the 3D XPoint subarray, that may potentially lead to errors in the results of TMVM. Let $G_{x}$ and $G_{y}$ be the conductances of the segments of $BL$s and $WL$s, respectively. The conductances for $WLT$ and $WLB$ are considered equal (both $G_y$) due to the symmetry and equal allocation of metal resources to $WLT$s and $WLB$s. We use $G_{i,j}$ to denote the conductance of PCM cell ($i,j$) at the top level, and $G_{O_{j}}$s to denote conductances of a column of PCM cells at the bottom level. In the worst case, each row performs an identical operation, and carries an equal current $I_{row}$. The total voltage drop to the last row is
\begin{equation}
\frac{I_{row}}{G_{y}}+\frac{2I_{row}}{G_{y}}+...+ \frac{N_{row}I_{row}}{G_{y}}=\frac{N_{row}(N_{row}+1)I_{row}}{2G_{y}}
\label{eq:voltage_drop}
\end{equation}
 where the first, second, and last terms on the left side of the equation are for voltage drops of $Segment_{N_{row}}$, $Segment_{N_{row}-1}$, and $Segment_{1}$, respectively. The voltage drop of the last row increases quadratically with the number of rows, and this causes a significant limit on the accuracy of the implementations~\cite{parasitics_importance_2020,CRAM_parasitics}. Hence, it is important to find the maximum allowable subarray size in which the voltage drop does not impair the electrical of implementations.

During the computation, the resistive network shown in Fig.~\ref{fig:parasitics_crk} can have different configurations based on the applied voltage to $WLT$s. For example, if $V_{2}\leftarrow$ {\em float}, then all $G_{2,j}$s and their connected parasitics must be removed from the equivalent circuit model of Fig.~\ref{fig:parasitics_crk}. To analyze parasitic effects, we consider the corner case for voltage drop, where only $V_{0}\leftarrow V_{DD}$ and the rest of the $V_{i}$s are floated, resulting in minimum equivalent conductance for inputs and wire parasitics. Moreover, for the corner case, we assume that inputs and outputs are located $N_{column}$ columns away from each other (the farthest possible distance). The value of all inputs assumed to be 1, and therefore, the current flows from the inputs of the TMVM computation must be sufficient enough to change the state of the output of the TMVM computation. An excessive voltage drop across the input and output cells causes a failure in the TMVM implementation discussed earlier.

\begin{figure}[ht!]
	\centering
	\includegraphics[width=9.0cm, height=2.8cm]{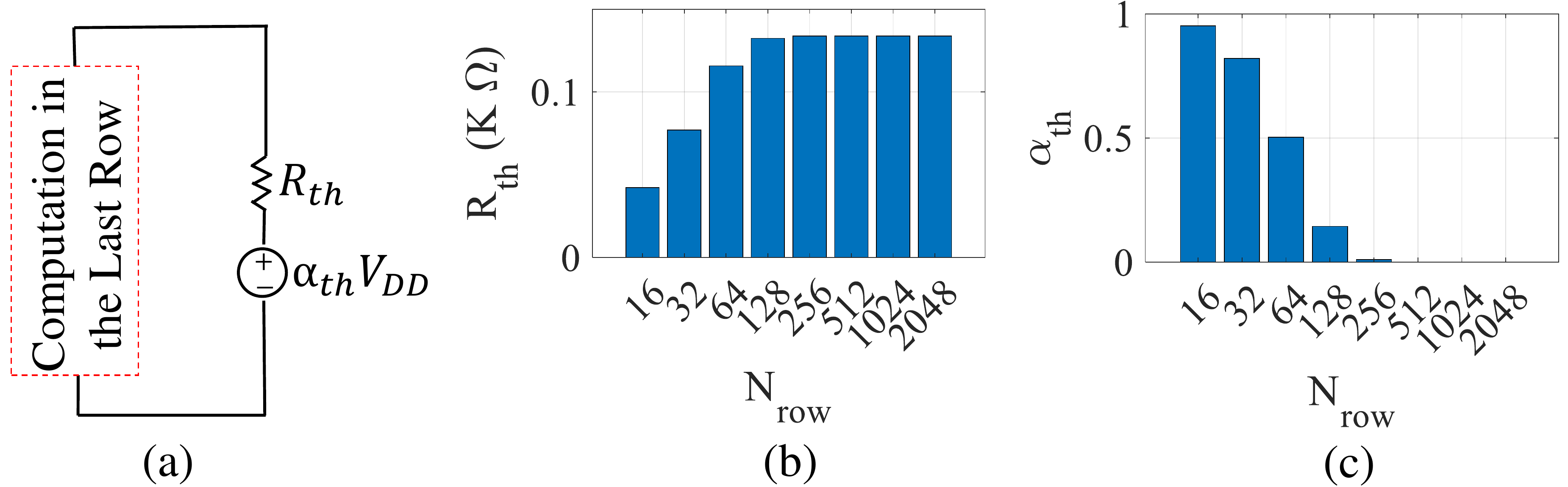}
	\caption{(a) Thevenin equivalents can be observed from the last row, (b) effects of $N_{row}$ on $R_{th}$, (c) and on $\alpha_{th}$.}
	\label{fig:Rth_ath_bar}
\end{figure}

The rows far from the drivers have larger parasitics between them and the driver. In particular, for the last row (farthest from the driver, see Fig.~\ref{fig:parasitics_crk}), the voltage drop is the worst. If the electrical correctness for the last row does not hold up, the implementations would be unacceptable. We observe the rest of the circuit from the last row and calculate (for the worst case) the Thevenin resistance ($R_{th}$) and Thevenin voltage ($V_{th}$) (see Fig.~\ref{fig:Rth_ath_bar}(a)). We define the Thevenin coefficient, $\alpha_{th} = \frac{V_{th}}{V_{DD}}$, and its value is between 0 and 1. Both $R_{th}$ and $\alpha_{th}$ can be obtained analytically using a recursive approach explained in Appendix~\ref{sec:appendix}. Both are functions of parameters such as $N_{row}$, $N_{column}$, PCM cell width ($W_{cell}$) and length ($L_{cell}$) as well as other parameters of PCM and wire devices. Fig.~\ref{fig:Rth_ath_bar}(b) and (c)  shows $R_{th}$ and $\alpha_{th}$ for different $N_{row}$ values. The configuration of lines are based on configuration 1 that will be discussed in Table~\ref{tbl:config_tbl} in the next Section.
\ignore{
 Increasing $N_{row}$ causes an increase in $R_{th}$ until it reaches a saturated level, and contrary, $\alpha_{th}$ decreases rapidly and become insignificant for large $N_{row}$s.
}

\begin{figure}[ht!]
	\centering
	\includegraphics[width=9cm, height=3.6cm]{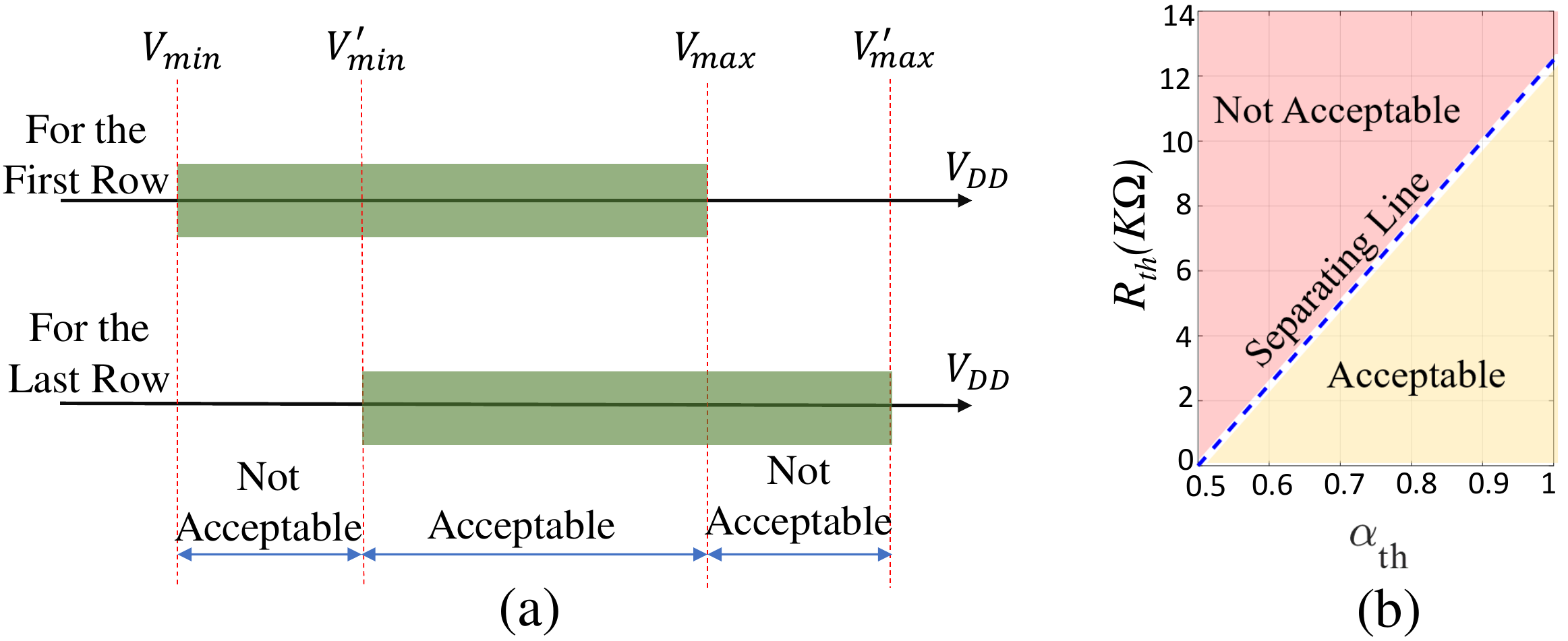}
	\caption{(a) Calculated voltage ranges for the first and last rows. (b) Acceptable and unacceptable regions in the ($\alpha_{th}, R_{th}$) plane.}
	\label{fig:R_th_alpha}
\end{figure}

There are negligible parasitics between the first row and the driver, and the voltage range that ensures accuracy of computing in the first row is closer to $[V_{min}, V_{max}]$ (discussed in Section~\ref{sec:realization}) than that of the last row. For the last row, values of $\alpha_{th}$ and $R_{th}$ are significantly affected by parasitics. Let us assume that the new voltage range ensures electrical correctness of the last row is $[V_{min}^{'}, V_{max}^{'}]$. The voltage ranges for the first row and last row are shown in Fig.~\ref{fig:R_th_alpha}(a). \ignore{The calculated voltage range of the last row is relatively larger than that of the first row, because a portion of the voltage is dropped on the parasitics.} We use the voltage ranges of first row and last row as two corner cases (with least and most voltage drops, respectively), and we find a voltage range the satisfy both corner cases; the obtained voltage range guarantees the electrical correctness for intermediate rows as well. The final acceptable voltage range is the overlap between two voltage ranges shown in Fig.~\ref{fig:R_th_alpha}(a), $[V_{min}^{'}, V_{max}]$, ensuring all of the rows from first row to the last row receiving the proper voltage.

 The noise margin ($NM$) in implementations is defined by 
\begin{equation}
NM=\frac{V_{max}-V_{min}^{'}}{V_{mid}}   
\label{eq:NM}
\end{equation}  
where $V_{mid}=(V_{max}+V_{min}^{'})/2$. Clearly, we desire $NM\geq0$. In Fig.~\ref{fig:R_th_alpha}(b), the acceptable and unacceptable regions in the ($\alpha_{th}, R_{th}$) plane is shown. The $NM$ on the separating line is 0, above it $NM$ is negative (unacceptable), and below it $NM$ is positive (acceptable). Our goal is to choose wire configurations so that the corresponding ($\alpha_{th}, R_{th}$) of the design falls into the acceptable region with maximum $NM$ possible.

\section{Results and Discussion}
\label{sec:results and Discussion}

\begin{figure}
	\centering
	\includegraphics[width=7cm, height=4.0cm]{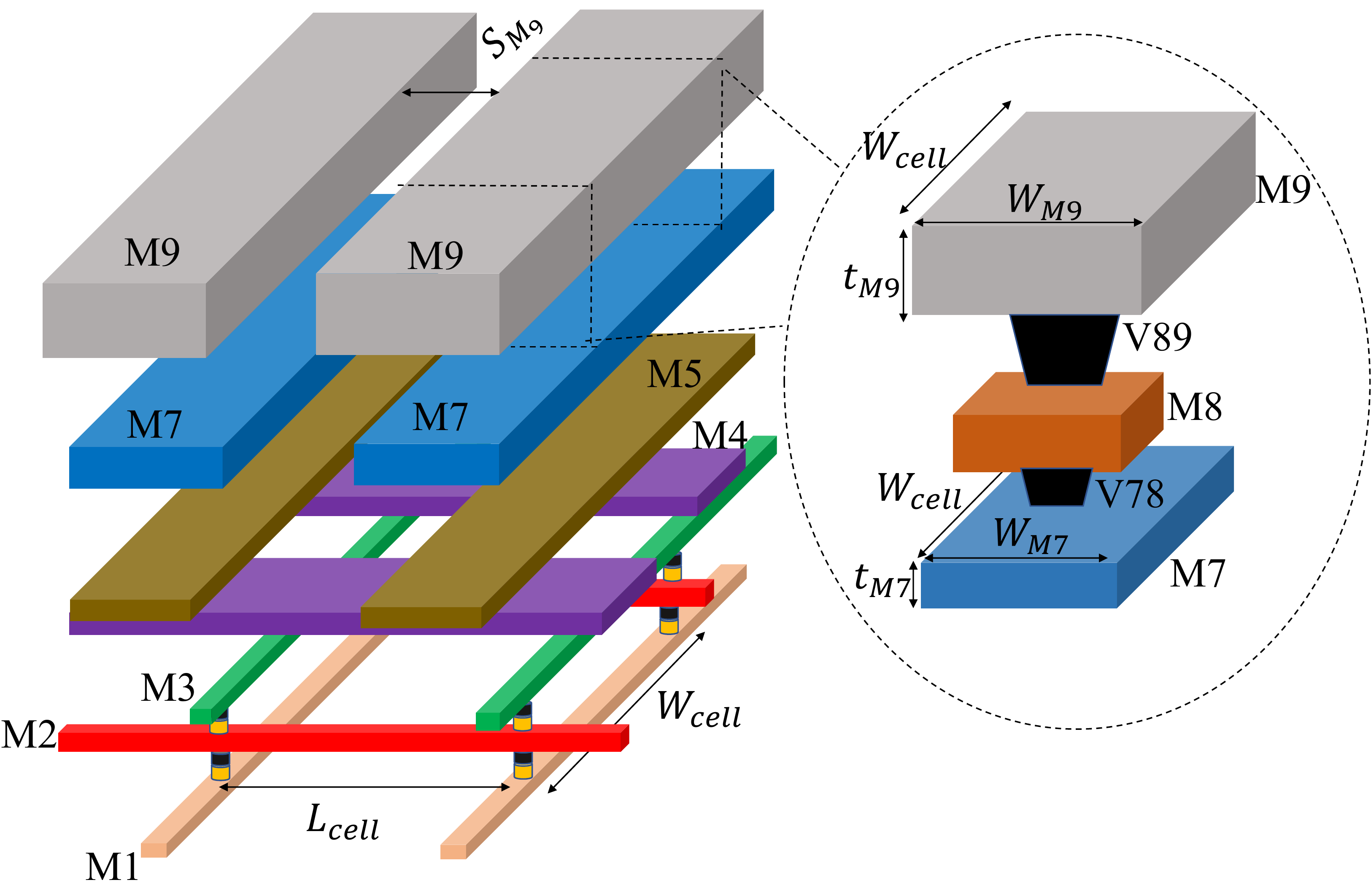}
	\caption{Multi-metal layer configuration can be utilized for the design of $WLT$s, $BL$s, and $WLB$s of 3D XPoint subarray.}
	\label{fig:metal}
\end{figure}

\begin{figure*}[ht!]
	\centering
	\includegraphics[width=18.5cm, height=4cm]{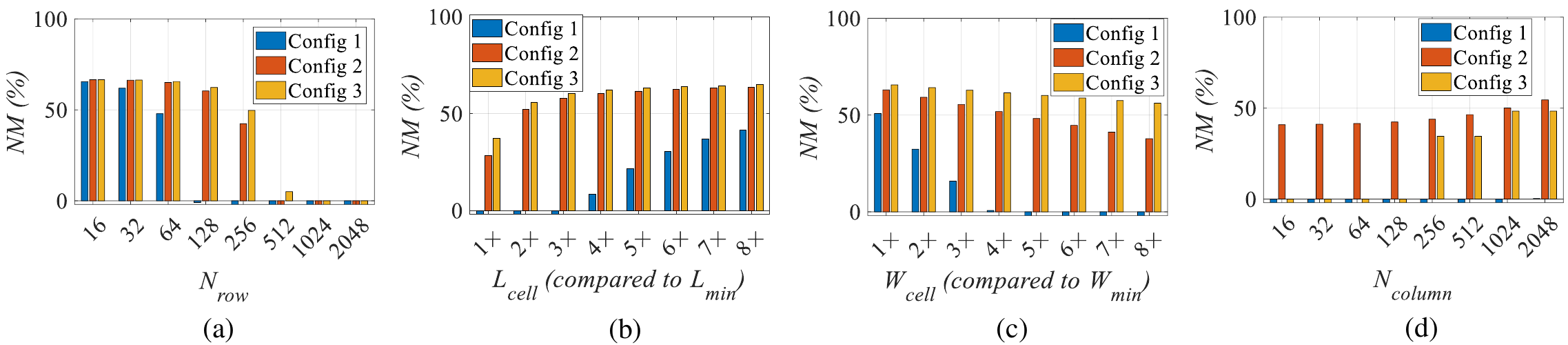}
	\caption{$NM$s of the three metal line configurations: (a) changing $N_{row}$ (while $N_{column}=128$, $L_{cell}=4 L_{min} $, and $W_{cell}=W_{min}$), (b) changing $L_{cell}$ (while $N_{column}=128$, $N_{row}=128 $, and $W_{cell}=W_{min}$), (c) changing $W_{cell}$ (while $N_{column}=128$, $N_{row}=64 $, and $L_{cell}=4L_{min}$), and (d) changing $N_{column}$ (while $N_{row}=256$, $L_{cell}=4L_{min}$, and $W_{cell}=W_{min}$).}
	\label{fig:config_NM}
\end{figure*}

\subsection{NM evaluation}

\begin{table}
	\centering 
	\caption{Different Configurations of Metal Lines in the 3D XPoint Subarray Based on ASAP7 Design Rules.}
	\includegraphics[width=8.0cm, height=2.0cm]{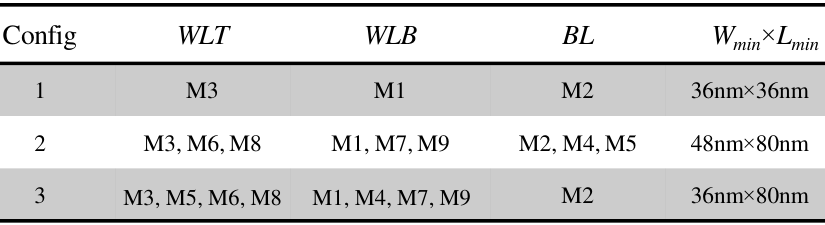}
	\label{tbl:config_tbl}
\end{table}

\noindent
To realistically analyze the effect of the parasitics, we assumed that metal layers in 3D XPoint are constructed based on ASAP7 design rules~\cite{ASAP7,Clark17} (see Fig.~\ref{fig:metal}). We can create different configurations for allocating metal lines to $WLT$s, $WLB$s, and $BL$s. Table~\ref{tbl:config_tbl} lists three possible configurations. In configuration~1, only M1, M2, and M3 (the first three metal lines) in ASAP7 are exploited for 3D XPoint, and they are allocated to $WLB$, $BL$, and $WLT$, respectively. For configurations~2 and 3, we assume that other than M1, M2, and M3, the other metal layers (M4 to M9) can also be allocated to the 3D XPoint lines. In configuration~2, we allocate M4 and M5 to the $BL$s, and M6 to M9 are allocated equally between $WLT$s and $WLB$s. In configuration~3, we assume that all metals from M4 to M9 are allocated equally between $WLT$s and $WLB$s; no extra top metal lines are allocated to $BL$s. We report the minimum cell width ($W_{min}$) and length ($L_{min}$) for each configuration based on the minimum required width of a line and space between adjacent lines in each layer. The values of parameters for metal lines and PCM devices are available in the Supplementary Material.

\noindent
{\bf $NM$ improves with increasing $N_{row}$}: Fig.~\ref{fig:config_NM}(a) shows $NM$s of different $N_{row}$ values. $NM$ is significantly sensitive to $N_{row}$. For $N_{row}$ as large as 2048, the implementations are not valid due to excessive voltage drop, and hence negative $NM$. Configuration 3 provides the best $NM$, because more metal resources dedicated to $WLT$ and $WLB$ causes smaller parasitics in the current path across rows.    

\noindent
{\bf $NM$ improves with increasing $L_{cell}$}: Fig.~\ref{fig:config_NM}(b) shows the $NM$s for different $L_{cell}$s (for each configuration, values are normalized to $L_{min}$ listed in Table~\ref{tbl:config_tbl}). By increasing $L_{cell}$, the width of the $WLT$s and $WLB$s increase, decreasing the parasitic resistances related to $WLT$s and $WLB$s.

\noindent
{\bf $NM$ decreases with increasing $W_{cell}$}: Fig.~\ref{fig:config_NM}(c) shows the $NM$s for different $W_{cell}$ (for each configuration, values are normalized to $W_{min}$ listed in Table~\ref{tbl:config_tbl}). By increasing $W_{cell}$, the length of the $WLT$s and $WLB$s increase, and consequently, parasitics related to $WLT$s and $WLB$s considerably increase. Therefore, for all cases, smaller $W_{cell}$ causes larger $NM$.     

\noindent
{\bf $NM$ remains unchanged with increasing $N_{column}$}: Fig.~\ref{fig:config_NM}(d) shows that the increase in $N_{column}$ does not affect $NM$ significantly. By increasing $N_{column}$, parasitics of $BL$s increase. However, since the $BL$ resistances are in series with those of PCM devices with orders of magnitude larger resistance, the increase in $BL$ resistance does not affect $NM$.

\subsection{Implementing NNs on the 3D XPoint substrate}

\noindent
We list the performance of various 3D XPoint subarrays of various sizes for the digit recognition of MNIST dataset in Table~\ref{tbl:result_tbl}. Each MNIST test image is scaled to 11$\times$11 as in \cite{MNIST_Kim}, a transformation that maintains 91\% recognition accuracy and reduces computation. We use configuration~3 in all cases, as it provides the best $NM$ among all alternatives. For the smallest subarray with size $64\times128$, $NM$ is the maximum among all cases. For the largest subarray with size $1024\times 1024$, we increase $L_{cell}$ by 2.6$\times$ (compare to that of $64\times128$ subarray) to decrease the parasitics of lines. Consequently,  we  achieve acceptable $NM$ of $34.5\%$. With this relatively large subarray, we have more parallelism that allow to process a larger number of MNIST images in each computational step, reducing the total execution time (17 $\times$ faster than that of 64$\times$128 subarray). The energy per image is similar for all cases because the subarray sizes listed in Table~\ref{tbl:result_tbl} are large enough to allow fully processing an 11$\times$11 MNIST image locally without the need for extra data movement between subarrays or peripheral circuitry.

We analyze the energy and area for the implementation a multi-bit TMVM using two schemes that we introduced in Section~\ref{sec:more_complex}D. We listed the results in Table~\ref{tbl:result_tbl2}. As we increase the number of bits for the $G_{i,j}$s, the allocated area for both implementations increases. However, while for area-efficient scheme, the area increases linearly, for the low-power scheme, the area increases exponentially. The implementation energy in the low-power scheme slightly increases with increasing the number of bits, while for the area-efficient schemes, energy increases rapidly. For the area-efficient scheme, we do not list the energy and area values beyond 3 bits, because it requires applying a large voltage level ($>$5V) within the subarray, making the implementation infeasible and unrealistic.

\begin{table}[t]
	\centering 
	\caption{Evaluation of Different Subarray Sizes for Digit Recognition Application.}
	\includegraphics[width=9.0cm, height=3.6cm]{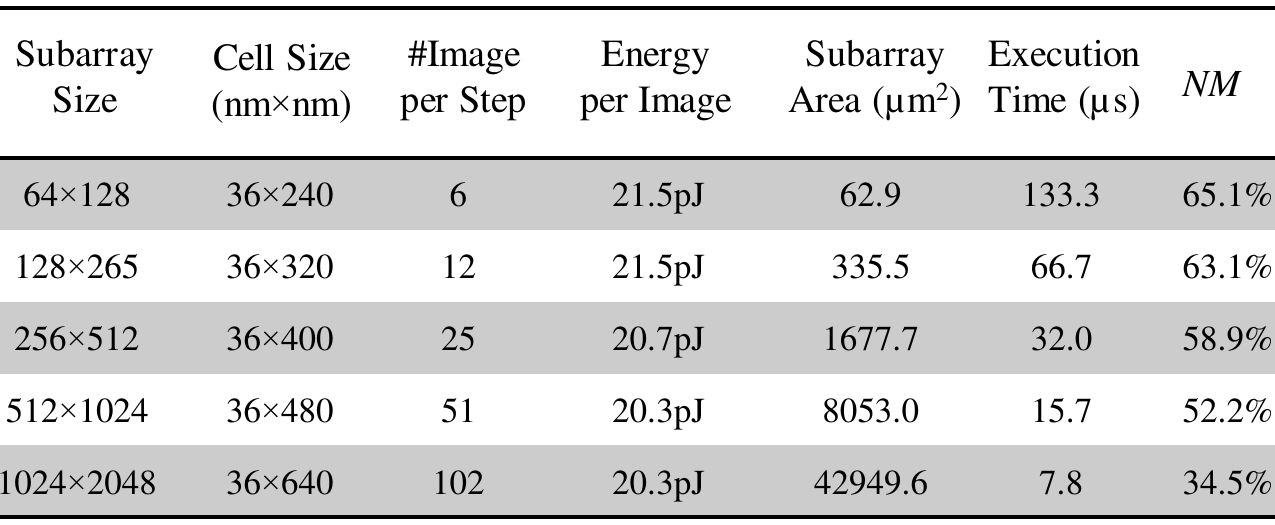}
	\label{tbl:result_tbl}
\end{table}

\begin{table}[t]
	\centering 
	\caption{Evaluation of Implementation Energy and Area for Multi-Bit TMVM using Area Efficient and Low Power Schemes.}
	\includegraphics[width=9.0cm, height=3.2cm]{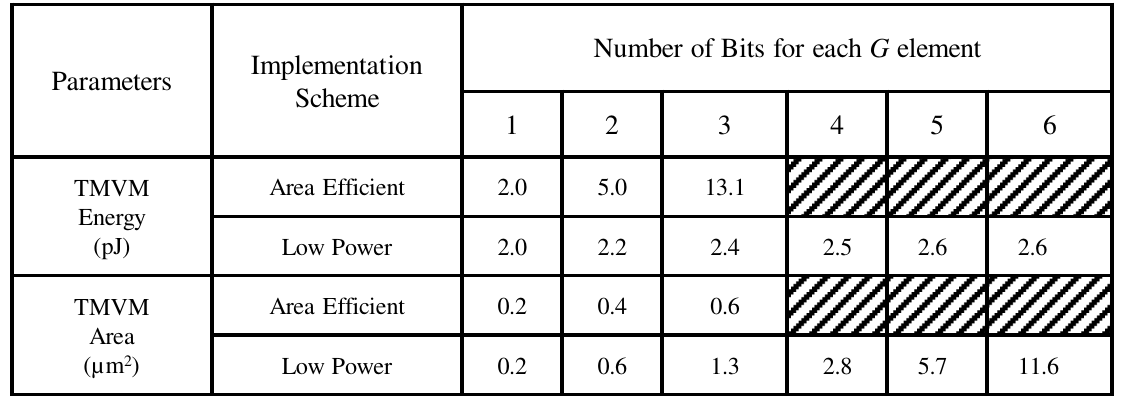}
	\label{tbl:result_tbl2}
\end{table}

\ignore{
\begin{table}
	\centring 
	\caption{Performance Comparison of a NMP System and 3D XPoint Accelerator}
	\includegraphics[width=9.0cm, height=3cm]{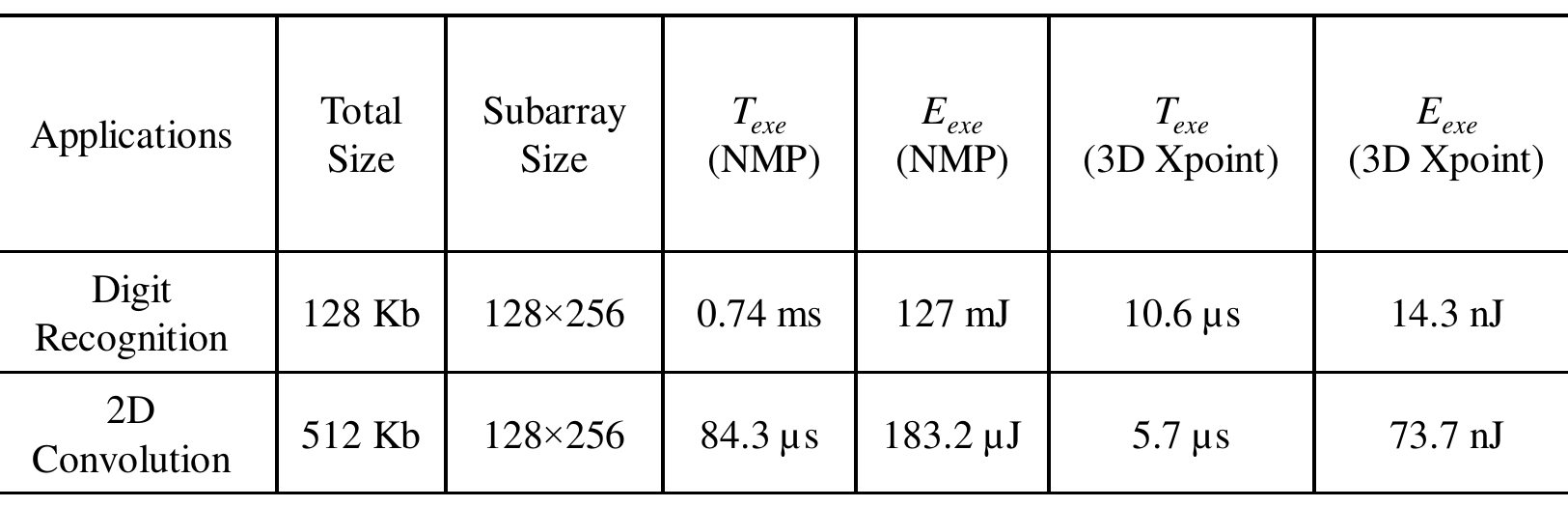}
	\label{tbl:eval_tbl}
\end{table}
}

\section{Conclusion}
\label{sec:conclusion}

\noindent
We have presented methods for the implementations of TMVM,
NN, and 2D convolution on 3D XPoint. To ensure the accuracy
of the implementations, we considered wire parasitics in
our implementations. We have demonstrated that interconnect
parasitics have a significant effect on the implementations
performance and have developed a comprehensive model
for analyzing this impact. Using this methodology, we have
developed guidelines for the 3D XPoint Subarray size and
configurations based on ASAP7 technology design rules. We
used different size 3D XPoint subarrays for digit recognition of MNIST dataset. Using the our methodology methodology, we design a relatively  large subarray of 2 Mb with acceptable $NM$ of 34.5$\%$, providing the opportunity for processing more images per step without any energy overhead. 

\ignore{
\begin{figure*}[!ht]
	\centering
	\subfigure[]{\includegraphics[width=2.2in, height=7.4cm]{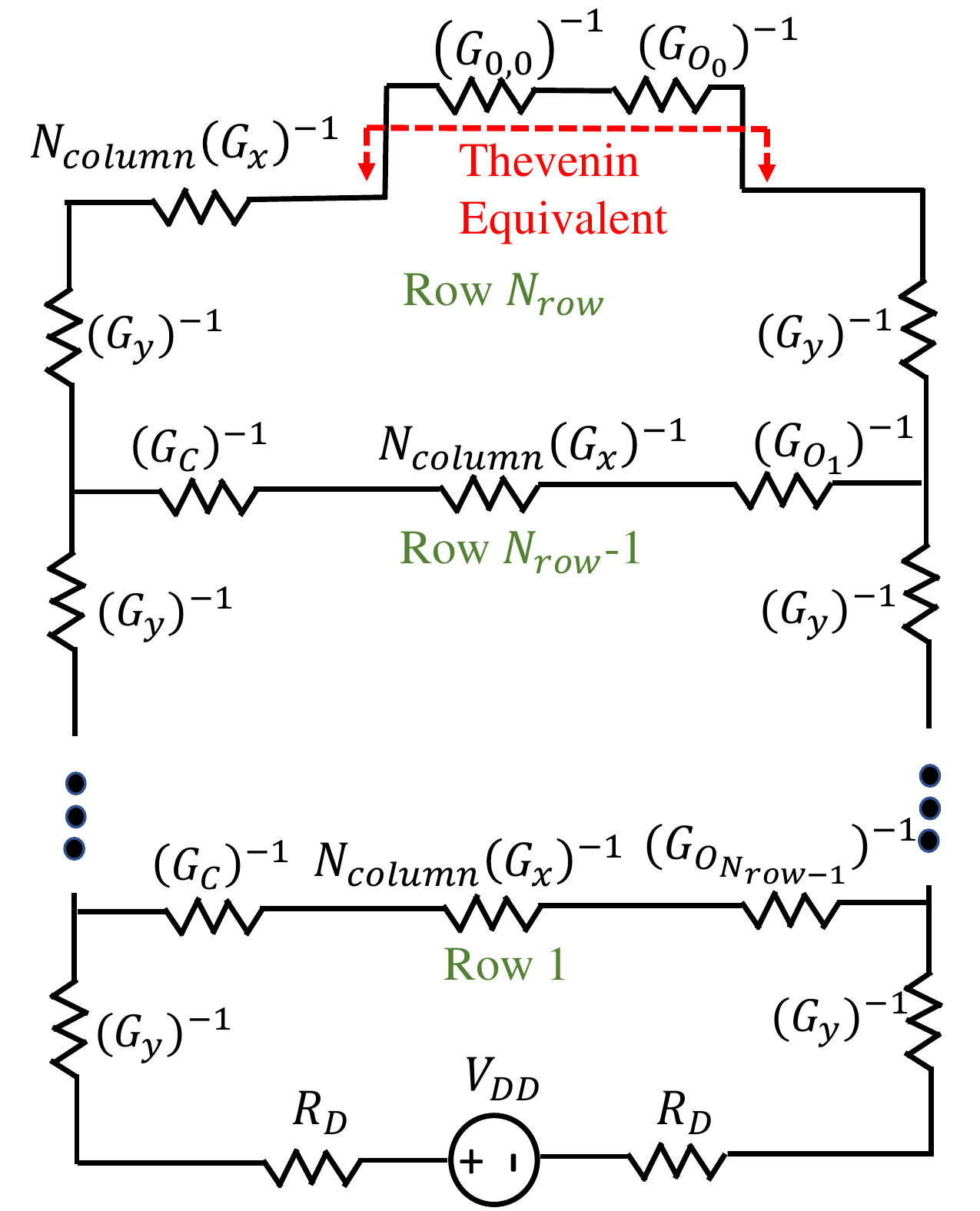}}
	\subfigure[]{\includegraphics[width=2.2in, height=7.4cm]{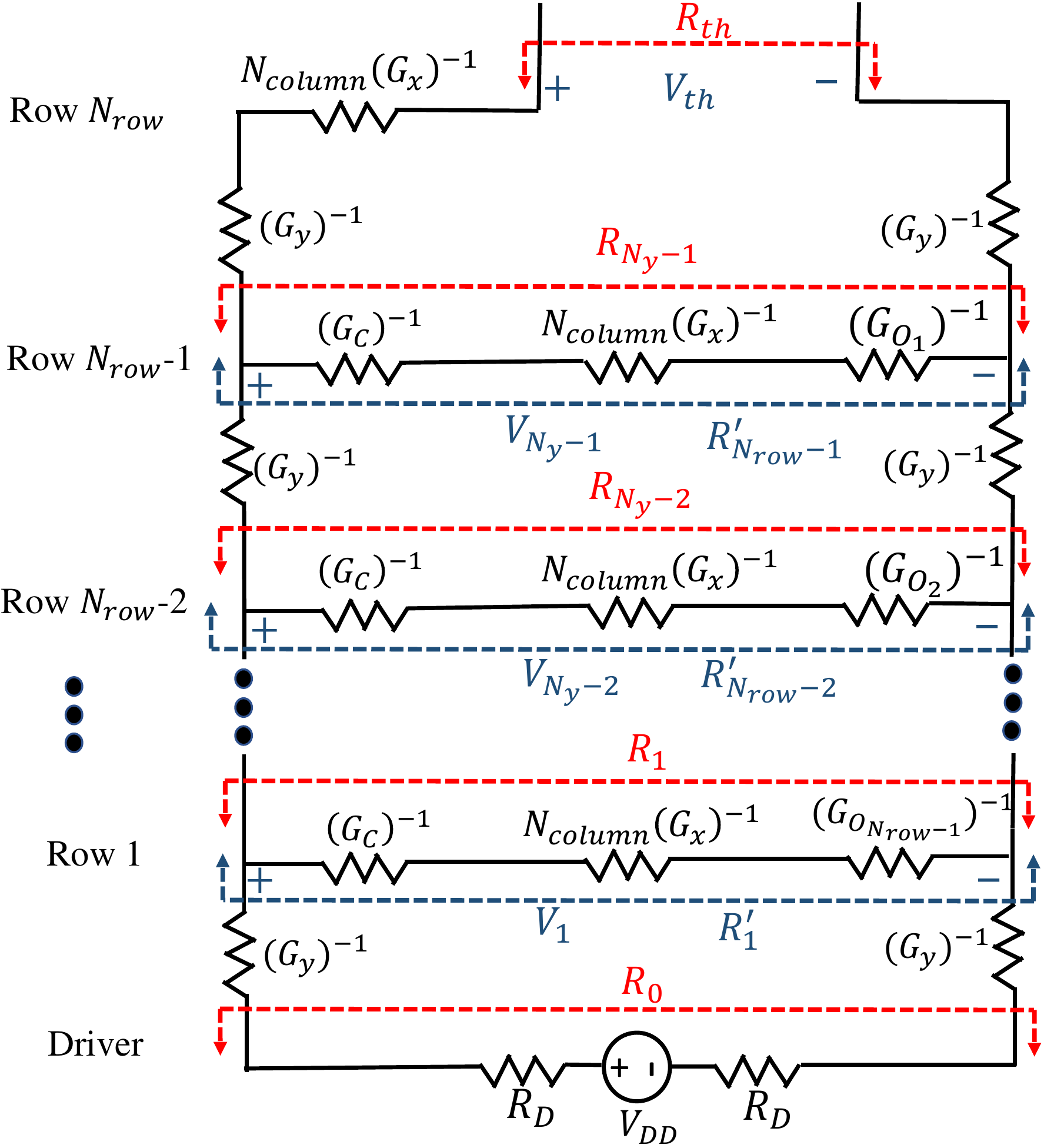}}
	\subfigure[]{\includegraphics[width=2.2in, height=7.4cm]{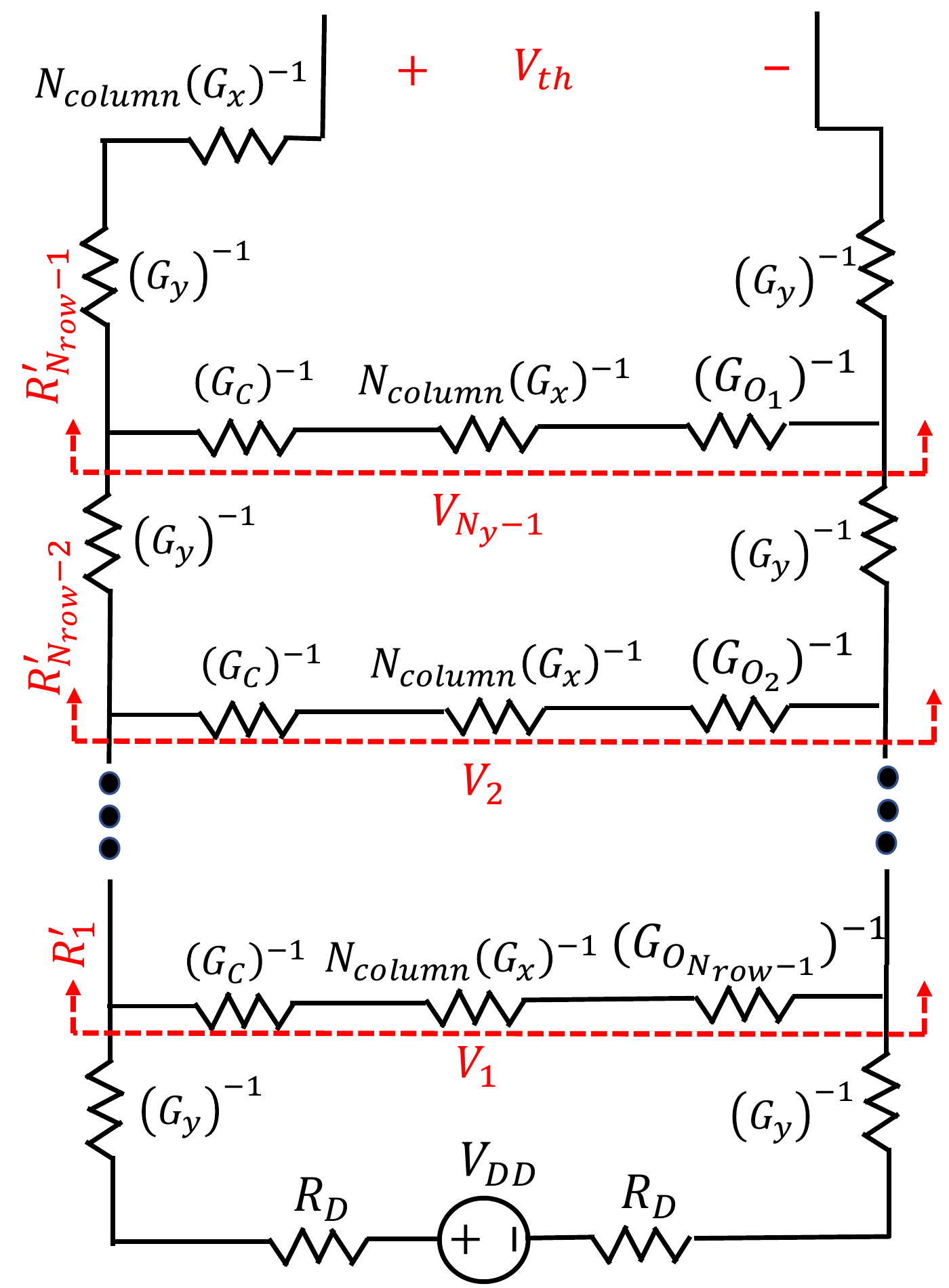}}
	\caption{(a) The circuit model for the implementation of TMVM in the worst case scenario, showing the observation point for calculating Thevenin equivalents; Notations used in the chain of rows for defining the (b) Thevenin resistance ($R_{th}$),  and (c) Thevenin voltage ($V_{th}$).}
	\label{fig:path_model_1in}
\end{figure*}
}

\bibliographystyle{misc/ieeetr2}
\bibliography{misc/bibfile}

\appendix
\label{sec:appendix}

\begin{figure}
	\centering
	\includegraphics[width=7.4cm, height=7.5cm]{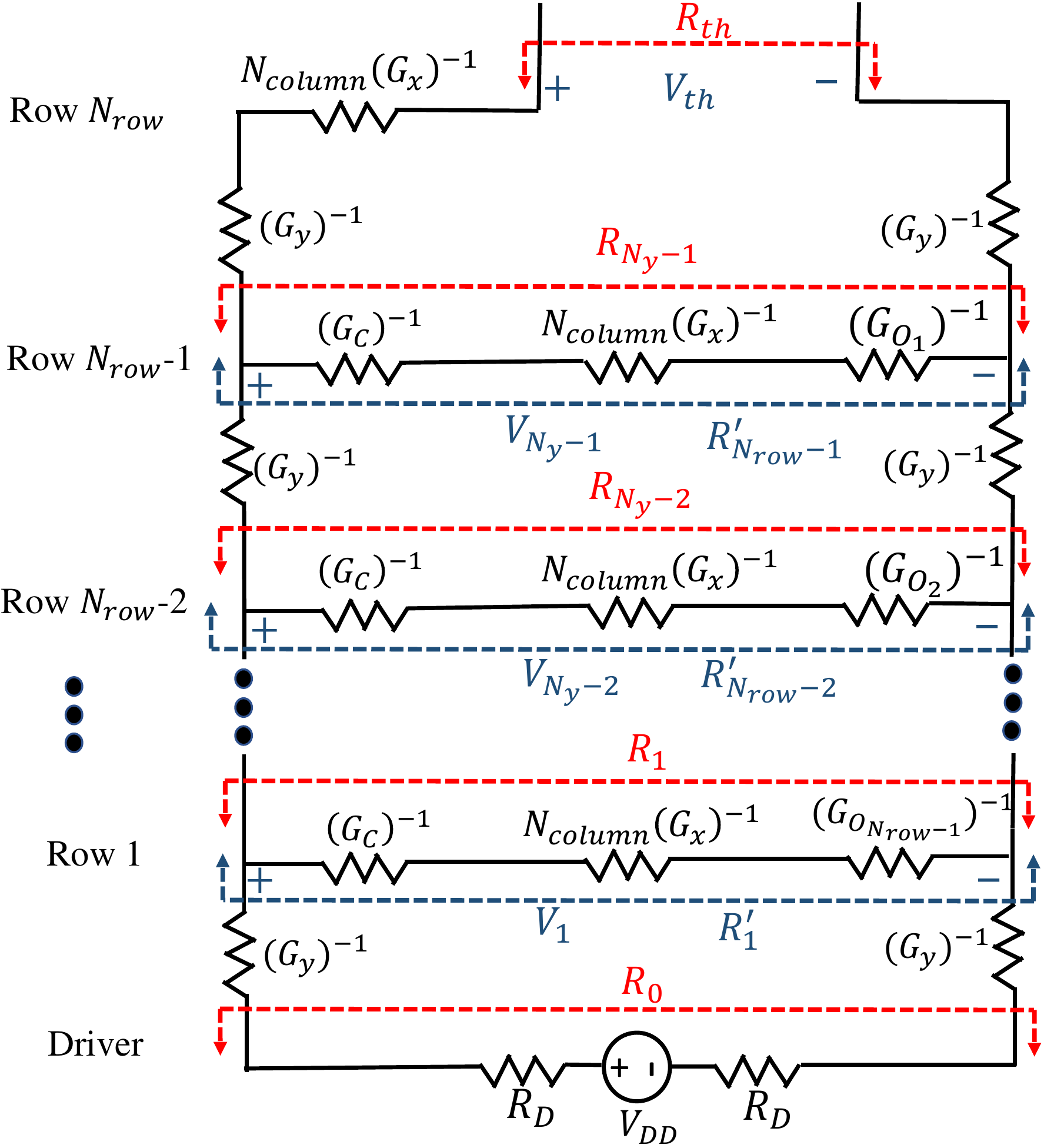}
	\caption{Notations used for calculating Thevenin resistance ($R_{th}$)  and Thevenin voltage ($V_{th}$) shown on  the circuit model for the implementation of TMVM in the worst case scenario.}  
	\label{fig:path_model_1in}
\end{figure}

\noindent
We derive recursive expressions for calculating $R_{th}$ and $V_{th}$ of a $\left(N_{row} \times N_{column}\right)$ subarray of 3D XPoint.

Within the footprint area of a cell $\left(W_{cell}\times L_{cell}\right)$, we define $G_{y}$ (representing the conductance for $WLT$ and $WLB$ segments) and $G_{x}$ (representing the conductance of $BL$ segment). Fig.~\ref{fig:path_model_1in} shows the equivalent simplified circuit  model for the implementation TMVM in the corner case. Row $i$ is separated from its predecessor by conductance $G_y$ at each end. The input cell is connected to the output cell, $N_{column}$ columns away. The conductances in the last row are rearranged to create a two-port structure consisting of the PCM conductances so that the rest of the network can be modeled using Thevenin Equivalents ($R_{th}$ and $V_{th}$).

For configuration 1 (listed in Table~\ref{tbl:config_tbl}), $G_{y}=G_{M_{1}}=G_{M_{3}}$ (assuming similar wire conductance for $WLT$s and $WLB$s) and $G_{x}=G_{M_{2}}$ in which the conductance, $G_{M_{k}}$, is given by:
$G_{M_{k}}^{-1}=\frac{\rho_{M_{k}}L_{M_{k}}}{t_{M_{k}}W_{M_{k}}}$,
where $\rho_{M_{k}}$, $L_{M_{k}}$, $t_{M_{k}}$, and $W_{M_{k}}$ are, respectively, the resistivity, length, thickness, and width in metal layer $k$ (see the Supplementary Material). For configurations 2 and 3, the equivalent conductance of the wire segment must be calculated based on the multi-metal layer configuration of a given segment. For example, in configuration 2, $G_{y}$ (representing a segment conductance of $WLT$) is obtained by $G_{y}= G_{M_3} + G_{M_6} + G_{M_8}$.

To calculate $R_{th}$ and $V_{th}$, we derive recursive expressions. For conciseness, we define the resistance, $R_{row_{i}}$, of row $i$ as:
\begin{equation}
R_{row_i} = N_{column}\left(G_{x}\right)^{-1}+\left(G_{C}\right)^{-1}+\left(G_{O_{N_{row}-i}}\right)^{-1}
\label{eq:R_rowi}
\end{equation}	
We can obtain $R_{th}$, using the notations in Fig.~\ref{fig:path_model_1in}, as:
\begin{equation}
R_{th} = 2\left(G_{y}\right)^{-1}+N_{column}\left(G_{x}\right)^{-1}+R_{N_{row}-1}
\label{eq:R_th_1in}
\end{equation}
where $R_{N_{row}-1}$ is calculated using the recursive expression:
\begin{equation}
R_i=\left(R_{row_{i}}\right)||\left(R_{i-1}+2\left(G_{y}\right)^{-1}\right)
\label{eq:R_i_1in}
\end{equation}
The base case corresponds to the driver row that precedes the first row, and is $R_0 = 2 R_D$, as seen in Fig.~\ref{fig:path_model_1in}. 

To compute $V_{th}$, as illustrated in Fig.~\ref{fig:path_model_1in}, we first compute the intermediate variable $R_{j}^\prime$, which corresponds to the effective downstream resistance (away from the source) seen from node $j$. The computation proceeds in a recursive fashion from the last row towards the first as:
\begin{equation}
R_{j-1}^\prime=\left(R_{row_{j-1}}\right)||\left(R_j^\prime+2\left(G_{y}\right)^{-1}\right)
\label{eq:R_prim_1in}
\end{equation}
with the base case $R_{N_{row}-1}^\prime = R_{row_{N_{row}-1}}$. Having computed $R_j^\prime$, we may now compute $V_{th} = V_{N_{row}}$, using a recursive computation on $V_i$:
\begin{equation}
V_j = \frac{R_j^\prime}{2\left(G_{y}\right)^{-1}+ R_j^\prime} V_{j-1}
\label{eq:V_j_1in}
\end{equation}
in which $2\leq j \leq N_{row}-1$ and the base case is:
\begin{equation}
V_1 = \frac{R_1^\prime}{R_1^\prime +2\left(G_{y}\right)^{-1}+ 2R_D} V_b
\label{eq:V_1_1in}
\end{equation}


\clearpage
\section*{Supplementary Material}
\label{sec:supplementary}

\subsection{PCM parameters}

The conductance values of parameters in a PCM cell listed in Table~\ref{tbl:parameters}. In this work, we adopt the RESET current ($I_{RESET}$) of 100$\mu$A with RESET Time ($t_{RESET}$) of 15ns, and SET time ($t_{SET}$) of 80ns with the assumption of SET current ($I_{SET}$) of 50$\mu$A ($=\frac{I_{RESET}}{2}$)~\cite{KIST_2019,KIST_2020}. 

	\begin{table}[!ht]
	\centering
	\caption{PCM cell parameters and values~\cite{KIST_2019,KIST_2020}}
	\label{tbl:parameters}
	\begin{tabular}{|c|c|c|}
		\hline
		Parameters & Description                                                                                    & Value                                                                                                               \\ \hline
		$G_{A}$    & \begin{tabular}[c]{@{}c@{}}PCM conductance in the \\ amorphous state\end{tabular}              & 660 n$\Omega^{-1}$                                                                                                  \\ \hline
		$G_{C}$    & \begin{tabular}[c]{@{}c@{}}PCM conductance in the \\ crystalline state\end{tabular}            & 160$\mu\Omega^{-1}$                                                                                                   \\ \hline
		$S_{1}$    & \begin{tabular}[c]{@{}c@{}}Voltage control switch \\ for OTS\end{tabular}                      & \begin{tabular}[c]{@{}c@{}}100n$\Omega^{-1}$(\textless{}0V) and \\ 10$\Omega^{-1}$(\textgreater{}0.3V)\end{tabular} \\ \hline
		$S_{2}$    & \begin{tabular}[c]{@{}c@{}}Voltage control switch for \\ PCM in crystalline state\end{tabular} & \begin{tabular}[c]{@{}c@{}}10$\Omega^{-1}$(\textless{}0.8V) and \\ 100n$\Omega^{-1}$(\textgreater{}1V)\end{tabular} \\ \hline
	\end{tabular}
\end{table}

\subsection{Interconnect specifications for ASAP7}

The interconnect specifications are listed in Table~\ref{tbl:ASAP7_metal},
which shows the metal thickness ($t_{M}$) and resistivity ($\rho_{M}$), the
minimum line spacing ($S_{min}$), minimum line width ($W_{min}$), and
Table~\ref{tbl:ASAP7_Vias}, which shows the via
parameters~\cite{ASAP7,Clark17}.

\begin{table}[th!]
\caption{Specification of Metal Layers in ASAP7~\cite{ASAP7,Clark17}}
\label{tbl:ASAP7_metal}
\centering
\setlength{\extrarowheight}{1mm}
\begin{tabular}{|c|c|c|c|c|c|}
\hline 
Metal & $t_{M}$ & $S_{min}$ & $W_{min}$   & $\rho_{M}$      \\ \hline
M1(V) & 36nm    & 18nm      & 18nm        & 43.2$\Omega$.nm \\ \hline
M2(H) & 36nm    & 18nm      & 18nm        & 43.2$\Omega$.nm \\ \hline
M3(V) & 36nm    & 18nm      & 18nm        & 43.2$\Omega$.nm \\ \hline
M4(H) & 48nm    & 24nm      & 24nm        & 36.9$\Omega$.nm \\ \hline
M5(V) & 48nm    & 24nm      & 24nm        & 36.9$\Omega$.nm \\ \hline
M6(H) & 64nm    & 32nm      & 32nm        & 32.0$\Omega$.nm \\ \hline
M7(V) & 64nm    & 32nm      & 32nm        & 32.0$\Omega$.nm \\ \hline
M8(H) & 80nm    & 40nm      & 40nm        & 28.8$\Omega$.nm \\ \hline
M9(V) & 80nm    & 40nm      & 40nm        & 28.8$\Omega$.nm \\ \hline
\end{tabular}
\end{table}

\begin{table}[ht!]
\caption{Specification of Vias in ASAP7~\cite{ASAP7,Clark17}}
\label{tbl:ASAP7_Vias}
\centering
\setlength{\extrarowheight}{1mm}
\begin{tabular}{|c|c|c|c|}
\hline
Via             &   $R_{V}$  &     Via Size     & Minimum Spacing \\ \hline
V12 (M1 and M2) & 17$\Omega$ & 18nm$\times$18nm &      18nm       \\ \hline
V23 (M2 and M3) & 17$\Omega$ & 18nm$\times$18nm &      18nm       \\ \hline
V34 (M3 and M4) & 17$\Omega$ & 18nm$\times$18nm &      18nm       \\ \hline
V45 (M4 and M5) & 12$\Omega$ & 24nm$\times$24nm &      33nm       \\ \hline
V56 (M5 and M6) & 12$\Omega$ & 24nm$\times$24nm &      33nm       \\ \hline
V67 (M6 and M7) & 8$\Omega$  & 32nm$\times$32nm &      45nm       \\ \hline
V78 (M7 and M8) & 8$\Omega$  & 32nm$\times$32nm &      45nm       \\ \hline
V89 (M8 and M9) & 6$\Omega$  & 40nm$\times$40nm &      57nm       \\ \hline
\end{tabular}
\end{table}

\subsection{Status of lines during communications between subarrays}
The status of 3D XPoint lines during the communications with each other is listed in Table~\ref{tbl:line_status}.

\begin{table}[h!]
	\centering
	\caption{Status of 3D XPoint Lines for Two Different Configurations}
	\begin{tabular}{c|c|c|c}
		\hline
		\multirow{2}{*}{Line}   & \multirow{2}{*}{Subarray} & \multicolumn{2}{c}{Configuration}                                                                                                                                                       \\ \cline{3-4} 
		&                           & BL-to-BL                                                                                    & BL-to-WLT                                                                                  \\ \hline
		\multirow{2}{*}{$WLT$s} & 1                         & $V_{i}$s are applied                                                                        & $V_{i}$ are applied                                                                        \\ \cline{2-4} 
		& 2                         & all float                                                                                   & all active                                                                                 \\ \hline
		\multirow{2}{*}{$BL$s}  & 1                         & all active                                                                                  & all active                                                                                 \\ \cline{2-4} 
		& 2                         & all active                                                                                  & \begin{tabular}[c]{@{}c@{}}all float -\{output row \\ connect to the ground\}\end{tabular} \\ \hline
		\multirow{2}{*}{$WLB$s} & 1                         & all float                                                                                   & all float                                                                                  \\ \cline{2-4} 
		& 2                         & \begin{tabular}[c]{@{}c@{}}all float-\{output column\\ connect to the ground\}\end{tabular} & all float                                                                                  \\ \hline
	\end{tabular}
	\label{tbl:line_status}
\end{table}

\subsection{Corner case circuit}
\begin{figure}[ht!]
	\centering
	\includegraphics[width=9cm, height=4.0cm]{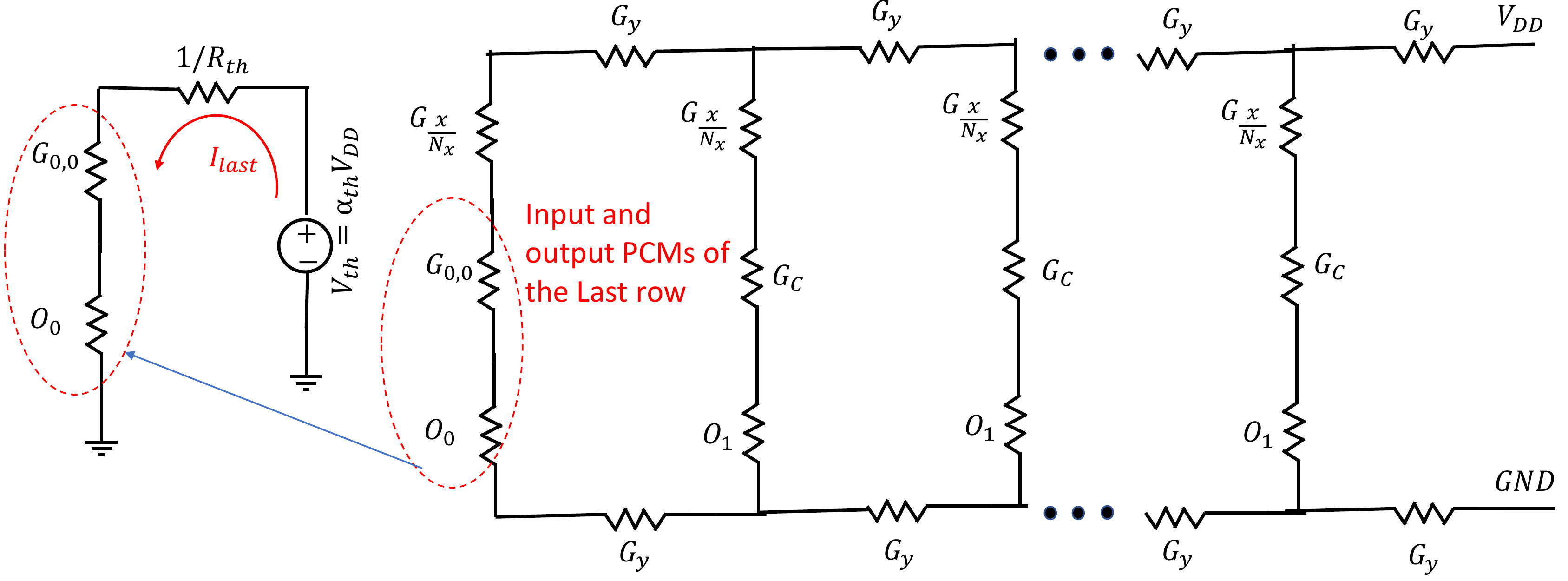}
	\caption{Reconfiguration and simplification of equivalent circuit model. }
	\label{fig:reconfigured_crk}
\end{figure}

The corner case circuit is shown in Fig.~\ref{fig:worst_case}. In the Appendix~\ref{sec:appendix}, we simplified the circuit even further and calculate the Thevenine equvalents observed from the last row. In Fig.~\ref{fig:reconfigured_crk}, we show the reconfiguration and simplification of circuit shown in Fig.~\ref{fig:worst_case}.

\begin{figure}[ht!]
	\centering
	\includegraphics[width=9cm, height=7.0cm]{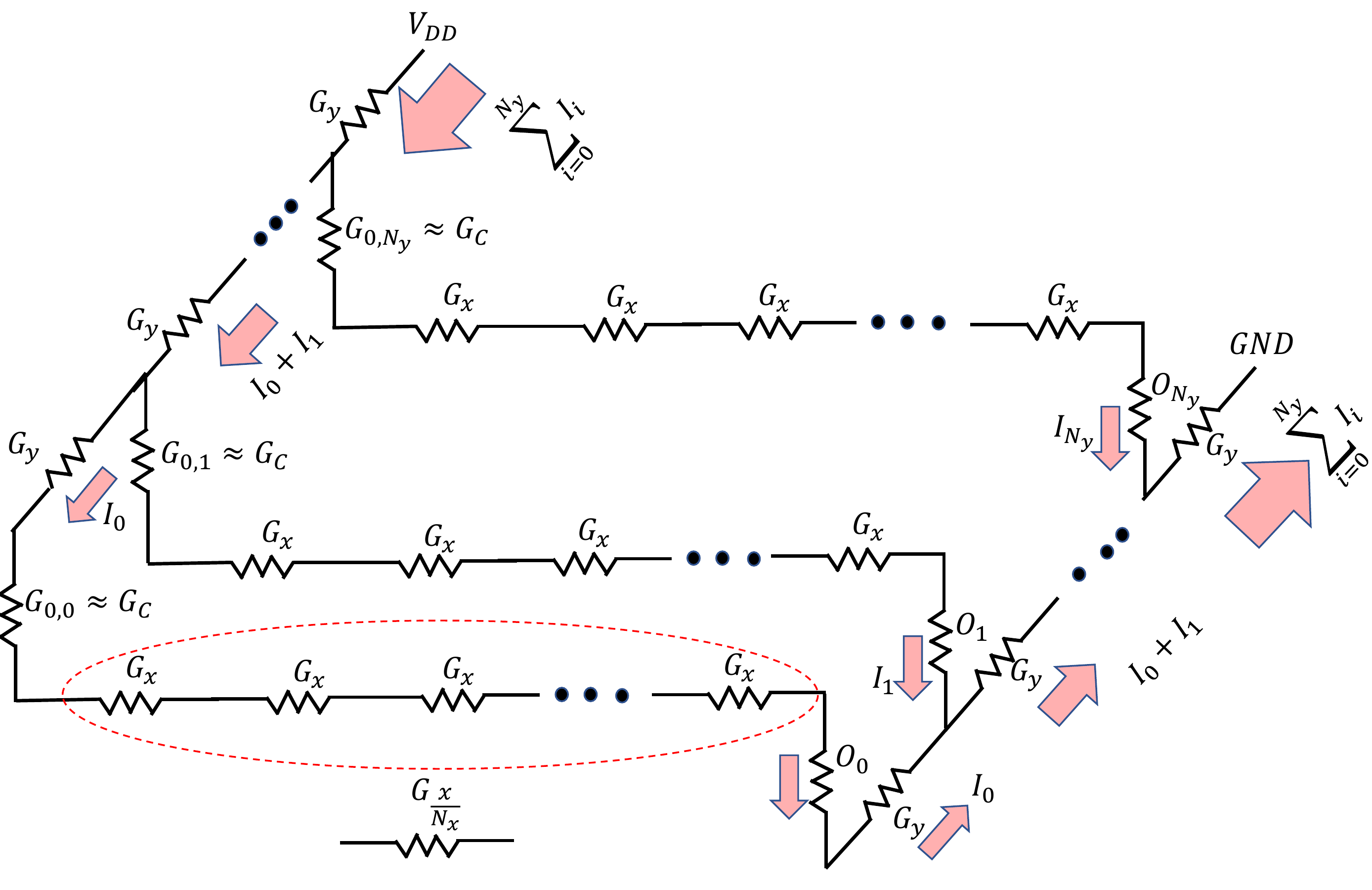}
	\caption{The equivalent circuit model for the worst case}
	\label{fig:worst_case}
\end{figure}

\end{document}